\documentclass{aa}
\pdfoutput=1  

\usepackage{graphicx}
\usepackage{txfonts}
\defcitealias{shalyapin14}{Paper I}
\usepackage{hyperref}
\begin{document} 

   \title{Gravitational lens system SDSS J1339+1310: microlensing factory and time delay}

   \author{L. J. Goicoechea\inst{1} \and V. N. Shalyapin\inst{1,2}}

   \institute{Departamento de F\'\i sica Moderna, Universidad de Cantabria, 
		 Avda. de Los Castros s/n, 39005 Santander, Spain\\
             \email{goicol@unican.es;vshal@ukr.net}
             \and
             Institute for Radiophysics and Electronics, National Academy of 
             Sciences of Ukraine, 12 Proskura St., 61085 Kharkov, Ukraine}


\abstract{We spectroscopically re-observed the gravitational lens system SDSS J1339+1310 using OSIRIS 
on the GTC. We also monitored the $r$-band variability of the two quasar images (A and B) with the LT 
over 143 epochs in the period 2009$-$2016. These new data in both the wavelength and time domains 
have confirmed that the system is an unusual microlensing factory. The C\,{\sc iv} emission line is 
remarkably microlensed, since the microlensing magnification of B relative to that for A, 
$\mu_{\rm{BA}}$, reaches a value of 1.4 ($\sim$ 0.4 mag) for its core. Moreover, the B image shows a 
red wing enhancement of C\,{\sc iv} flux (relative to A), and $\mu_{\rm{BA}}$ = 2 (0.75 mag) for the 
C\,{\sc iv} broad-line emission. Regarding the nuclear continuum, we find a chromatic behaviour of 
$\mu_{\rm{BA}}$, which roughly varies from $\sim$ 5 (1.75 mag) at 7000 \AA\ to $\sim$ 6 (1.95 mag) at 
4000 \AA. We also detect significant microlensing variability in the $r$ band, and this includes 
a number of microlensing events on timescales of 50$-$100 d. Fortunately, the presence of an 
intrinsic 0.7 mag dip in the light curves of A and B, permitted us to measure the time delay between 
both quasar images. This delay is $\Delta t_{\rm{AB}}$ = 47$^{+5}_{-6}$ d (1$\sigma$ confidence 
interval; A is leading), in good agreement with predictions of lens models.} 


   \maketitle
%
\clearpage

\section{Introduction}
\label{sec:intro} 

If there is a massive galaxy between a distant quasar and the Earth, the quasar is seen as a
multiple system consisting of several images \citep[e.g.][]{schneider92}. This gravitationally 
lensed quasar may also suffer microlensing effects by stellar mass objects in the lensing galaxy.
Quasar microlensing was firstly detected in 1989 \citep{irwin89,vander89}, and has become a powerful 
astrophysical tool in the current century \citep[e.g.][]{schneider06}. Microlensing particularly 
affects compact sources such as the X-ray emitting regions, the accretion disk or the innermost 
line emitting clouds, so microlensed quasars are being intensively used to probe the quasar 
structure from spectral studies \citep[e.g.][]{chartas04,richards04,pooley07,sluse07,bate08,floyd09,
blackburne11,mediavilla11,mosquera11,munoz11,jimenez12,motta12,sluse12,guerras13,braibant14,
jimenez14,rojas14,sluse15} and analyses of extrinsic variabilities \citep[e.g.][]{chartas02,
shalyapin02,goico03,kochanek04,gil06,morgan06,paraficz06,eigenbrod08,morgan08a,morgan08b,
poindexter08,chartas09,dai10,morgan10,poindexter10,chen11,sluse11,chartas12,chen12,morgan12,
hainline13,mosquera13,blackburne14,blackburne15,macleod15,mediavilla15a,mediavilla15b}. 

Unfortunately, many lensed quasars have only experienced weak microlensing effects in their spectra 
and light curves. Thus, it is important to identify the systems showing substantial spectral 
distortions due to microlensing, sharp microlensing variations or both phenomena. These microlensing 
factories are excellent tools for detailed interpretations, as well as the best targets for 
subsequent follow-up through suitable facilities \citep[e.g.][]{moskoc11}. For example, some systems 
exhibited strong chromatic microlensing in the optical continuum, with a maximum signal of $\sim$ 0.8 
mag at 5439 \AA\ \citep[HE 0047$-$1756 and SDSS J1155+6346;][]{rojas14}. A microlensing-induced 
distortion of the shape of several high-ionization emission lines was also unambiguously detected in 
SDSS J1004+4112 \citep[e.g.][]{richards04,motta12}. Additionally, the Einstein Cross (QSO 2237+0305) 
displayed prominent microlensing (extrinsic) events in its $V$-band light curves over the final years 
of the past century \citep{wozniak00,udalski06}, and these $\sim$ 0.7 mag variations on hundreds of 
days strongly stimulated interpretation tasks and further observations (see above). 

Lensed quasars are not only powerful laboratories of the quasar structure, but are also often 
used as a cosmological probe \citep[e.g.][]{refsdal64,schneider06}. For that purpose, the time delay 
between pairs of lensed images must be measured to high accuracy, requiring us to disentangle 
extrinsic and intrinsic variability. Intrinsic variations in a lensed quasar appear in two 
given images at different observing times, and a measurement of the time delay between these lensed 
images can be used to estimate the current expansion rate of the Universe (the so-called Hubble 
constant) and other cosmological parameters \citep[e.g.][]{oguri07,suyu10,suyu13,sereno14,rathna15}. 
The time delay is also sensitive to the distribution of the lensing mass \citep[e.g.][]{refsdal64,
schneider06,goico10}, so time delay measurements are turning in critical data for cosmology and 
extragalactic astrophysics. However, lensed quasars with appreciable extrinsic variations in their 
optical light curves may be a challenge for time delay determinations, since, in general, the 
extrinsic variability should be modelled by microlensing simulations or appropriate functions 
\citep[e.g.][and references therein]{tewes13a}. 

Various techniques have been developed to deblend the intrinsic signal from the 
microlensing variability. However, as seems intuitive, if the intrinsic and extrinsic variations have 
similar timescales and amplitudes, it is not possible to fairly distinguish between both signals and 
accurately determine the time delay of a lens system 
\citep[e.g. Q J0158$-$4325;][]{morgan08a,morgan12}. 
Despite the presence of clear extrinsic variability in HE 1104$-$1805, the time delay between its two 
images was measured to 4\% precision \citep[1$\sigma$ confidence interval;][]{ofek03}. \citet{ofek03} 
modelled a long-term microlensing gradient and analised the influence of short-timescale microlensing 
events (having a mean amplitude of $\sim$ 0.07 mag and a duration of approximately 1 month) in the 
delay estimation. Using additional data, and Legendre polynomials for describing the intrinsic and 
extrinsic variabilities, the relative uncertainty in the time delay measurement was decreased to 2\% 
\citep{poindexter07}. For this system (HE 1104$-$1805), \citet{morgan08a} also re-estimate the delay 
by modelling the extrinsic fluctuations through microlensing simulations. They obtained an 1$\sigma$ 
confidence interval in very good agreement with the Ofek \& Maoz's measurement. More recently, 
\citet{tewes13b} and \citet{rathna13} modelled 
the long-timescale extrinsic variations of two lens systems (RX J1131$-$1231 and SDSS J1001+5027) by 
free-knot splines (among other techniques), incorporating short-timescale correlated noise in their 
analyses. They measured the longest delays with $\leq$ 2\% precision. \citet{hainline13} also 
modelled the microlensing fluctuations in SBS 0909+532 by intensive simulations. The light curves of
this double quasar included an intrinsic deep dip and significant extrinsic variability on different 
timescales, allowing the authors to estimate the 50-d delay with 6\% precision.

As a part of the road map to deeply analyse the system \object{SDSS J1339+1310}, this paper is 
mainly dedicated to characterising the microlensing signal in the wavelength and time domains, 
as well as to measure the time delay. 
The lens system \object{SDSS J1339+1310} consists of two quasar images (A and B) at the same redshift
$z_{\rm{s}}$ = 2.231 and separated by 1\farcs70, as well as a lensing galaxy (G) at $z_{\rm{l}}$ = 
0.609 and located 0\farcs63 from the B image \citep{inada09,shalyapin14}. While A and B are optically 
bright ($r \sim$ 18$-$19 mag), the galaxy G has an $r$-band magnitude of about 20.5. Spectra of the 
lens system were obtained in 2013 using the OSIRIS R500R grism on the 10.4 m Gran Telescopio Canarias 
(GTC). These GTC-OSIRIS spectra allowed us to measure the redshift of G and 
obtain constraints on the macrolens and extinction parameters \citep[][henceforth Paper I]{shalyapin14}. 
We also found evidence for strong chromatic microlensing in its optical continuum (reaching a maximum 
signal of $\sim$ 1.5$-$1.7 mag at $\sim$ 5000 \AA), and were able to predict a time delay between A 
and B of $\sim$ 40$-$50 d (A is leading). The presence of sharp extrinsic events in preliminary light
curves with the 2.0 m Liverpool Telescope (LT) precluded direct measurement of the delay. 

In Sect.~\ref{sec:spec}, we present new high-quality GTC (optical) spectra of A, B and G in 2014. In 
Sect.~\ref{sec:microspec}, we discuss widely the several spectral distortions caused by microlensing. 
In Sect.~\ref{sec:lcur}, we also present $r$-band LT light curves of A and B in 2009, 2012$-$2015 and 
early 2016. In Sect.~\ref{sec:delmicvar}, these curves are used to estimate the time delay and the 
$r$-band microlensing variability. Our conclusions are summarised in Sect.~\ref{sec:end}. We also
include two appendices to address some specific issues. 

\section{Spatially resolved spectroscopy}
\label{sec:spec}

To improve the spectroscopic information of \object{SDSS J1339+1310}, 3570 (3$\times$1190) s 
GTC-OSIRIS exposures of the lens system were taken on 27 March 2014 (R500R grism) and 20 May 2014 
(R500B grism). The slit was oriented along the line joining both quasar images A and B, and the slit 
width was 1\farcs23 ($\sim$ 5 pixel). We used IRAF\footnote{IRAF is distributed by the National 
Optical Astronomy Observatory, which is operated by the Association of Universities for Research in 
Astronomy (AURA) under cooperative agreement with the National Science Foundation. This software is 
available at \url{http://iraf.noao.edu/}} packages to perform usual data reductions, and the raw and 
reduced spectral frames in FITS format are publicly available at the GTC 
archive\footnote{\url{http://gtc.sdc.cab.inta-csic.es/gtc/index.jsp}}. The new $\sim$ 1 h exposures 
are longer than the previous one with the R500R grism \citepalias{shalyapin14}. The new data also 
cover a broader wavelength range of 3600$-$9260 \AA\ \citep[the fringing and second-order 
contamination become relevant from about 9300 \AA\ redward; see sections 6.3 and 6.7 
of][]{cabrera14}. Additionally, the seeing conditions in 2014 were better than those in 2013, since 
we estimated seeing values of 0\farcs8, 0\farcs9 and 1\farcs1 for the three dithered sub-exposures 
with the red grism (7000 \AA), and 0\farcs9, 0\farcs9 and 1\farcs0 for the sub-exposures with the 
blue grism (4745 \AA). The spectral resolution of the two grisms was evaluated from the widths 
of the bright sky line at 5557 \AA, resulting in resolving powers of 290 ($FWHM$ = 19.2 \AA; red 
grism R500R) and 367 ($FWHM$ = 15.2 \AA; blue grism R500B). 

Before presenting the extraction of spectra for the three sources (A, B and G), we want to 
briefly discuss the differential atmospheric refraction (DAR) in our observations. This atmospheric 
phenomenon may cause a substantial chromatic offset of sources across the slit (transverse 
direction), leading to undesirable spectral artefacts \citep{filippenko82}. The amplitude of the 
wavelength-dependent transverse offset is related to the airmass and the slit position angle with 
respect to the zenith. Thus, the higher is the airmass and the farther the zenith is oriented the 
slit, the larger is the amplitude of the transverse offset. For the three GTC-OSIRIS sub-exposures 
with the red grism, the values of the airmass and the slit position angle were (1.223, 1.294, 1.383) 
and ($-$1\fdg8, $-$4\fdg0, $-$5\fdg6), respectively. For the blue grism sub-exposures, the airmass 
was lower (1.056, 1.076, 1.105), while the position angle was larger (24\fdg9, 15\fdg2, 8\fdg2). From 
these data, considering on-axis quasar images at 6200 \AA\ (acquisition frames were taken in the SDSS 
$r$ passband), the maximum transverse shifts of sources are 0\farcs05 (red grism) and 0\farcs2 (blue 
grism), and significant spectral distortions induced by DAR are not expected. The maximum spectral 
deviation by slit loss occurs on the blue edge of the blue grism, where a transverse shift of $\sim$ 
0\farcs2 induces a deviation of a few percent. This artefact is removed when calculating the spectral 
ratio $B/A$ in the bluest region. The spectral ratio $B/A$ is essentially free from DAR artefacts, 
because the two involved spectra are equally affected (same grism) or weakly affected (region of 
overlap between both grisms at 4850$-$7200 \AA) by DAR. 
 
The extraction of individual spectra of A, B and G is not a so direct task, and we followed the 
procedure that was presented and exhaustively discussed in \citetalias{shalyapin14}. Our technique is 
a variant of the approach by \citet{sluse07} to produce spectra of sources in a crowded region. 
We initially modelled the lens system as a 2D light distribution including two point-like sources (A 
and B) and a de Vaucouleurs profile (G). This ideal model relied on the astro-photometric constraints 
in the last column of Table 1 of \citetalias{shalyapin14}. The initial light distribution was then 
convolved with a 2D Moffat function, masked with the slit transmission and integrated across the 
slit. The final (realistic) 1D model at each wavelength depends on the position of A, the shape of 
the Moffat profile, and the fluxes of A, B and G. After some iterations (fits to the 1D data for each 
wavelength bin), we extracted the instrumental fluxes of the three sources throughout the entire 
spectral range of each individual frame (sub-exposure). The basic response curves (red and blue 
grisms) to calibrate the flux scales were derived from spectroscopic observations of the standard 
stars Hilt600 and GD153. Each star was modelled as a 2D Moffat distribution at each wavelength. These 
distributions were masked with the slit transmission, integrated across the slit and compared to the 
1D observed profiles. The basic responses were then corrected by airmass differences to get a 
response curve for each individual frame of the lens system. In a last step, the three instrumental 
spectra of each source for each grism were calibrated in flux and combined into a single spectral 
energy distribution.

\begin{figure}
\centering
\includegraphics[width=9cm]{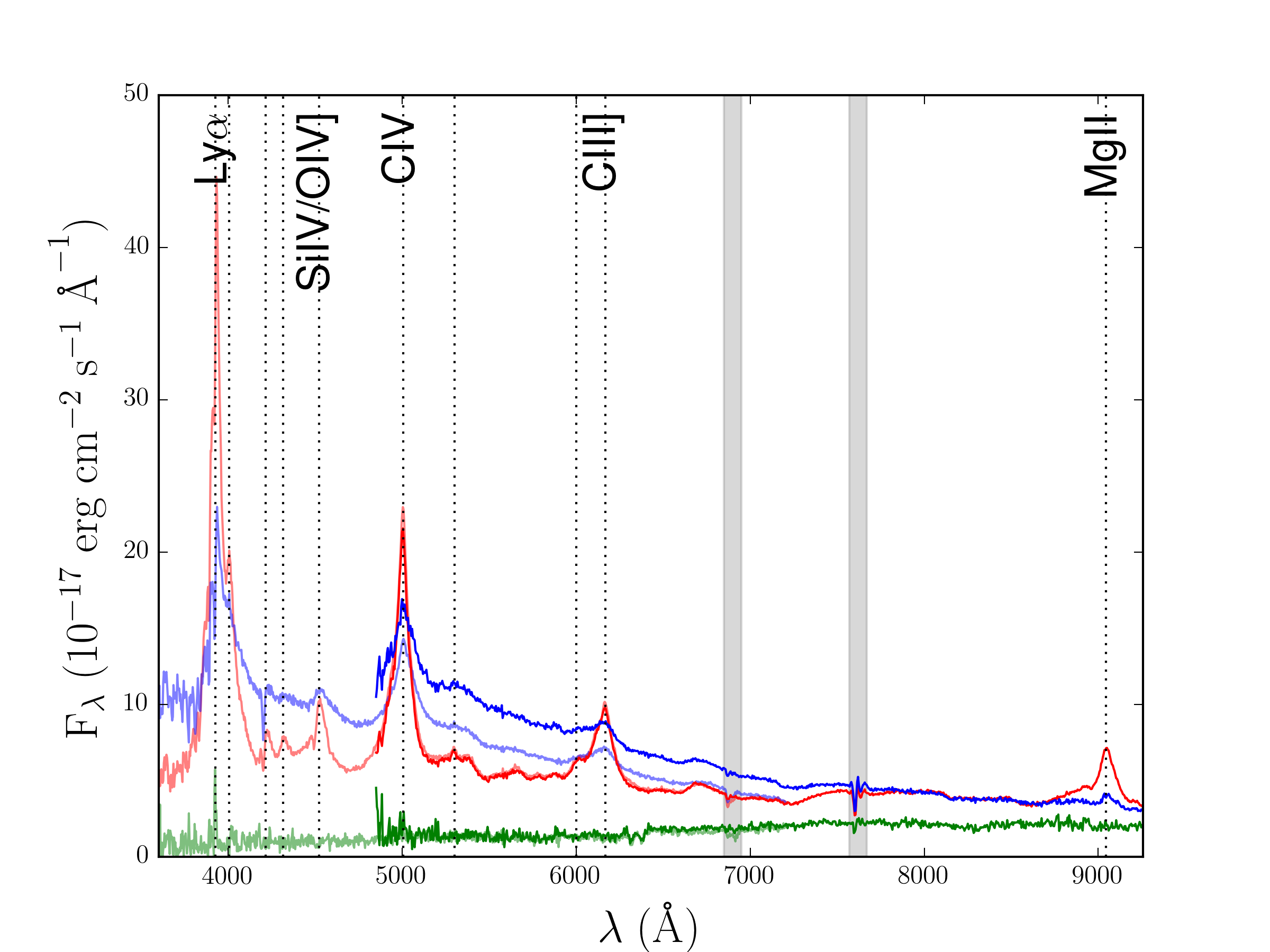}
\caption{GTC-OSIRIS spectra of SDSS J1339+1310ABG in 2014. The fluxes from the R500R (27 March) 
and R500B (20 May) grism data are depicted with bright and light colours, respectively (A in 
red, B in blue and G in green). Vertical dotted lines indicate emission lines at $z_{\rm{s}}$ = 
2.231, while grey highlighted regions are associated with atmospheric artefacts.}
\label{fig:specABG}
\end{figure}

The final spectra of A, B and G are available in tabular format at the CDS\footnote{See also 
\url{http://grupos.unican.es/glendama/q1339.htm}}: Table 1 includes wavelengths in \AA\ (Col. 1), and 
fluxes of A, B and G in 10$^{-17}$ erg cm$^{-2}$ s$^{-1}$ \AA$^{-1}$ (Cols. 2, 3 and 4) from the red
grism data, while Table 2 is structured in the same manner as Table 1, but incorporating the spectra 
associated with the blue grism. The results from the observations with both grisms are also plotted 
in Fig.~\ref{fig:specABG}. In this figure, we show the spectra of A (red), B (blue) and G (green). 
Bright colours trace the R500R spectra at the first observing epoch, whereas light colours represent 
the R500B spectra at the second epoch, $\sim$ 50 d later. As expected, the new quasar spectra are 
much less noisy than those in \citetalias{shalyapin14}, and incorporate many emission lines (vertical 
dotted lines). Here, in Sec.~\ref{sec:microspec}, we focus on the five resolved prominent features: 
Ly$\alpha$, Si\,{\sc iv}/O\,{\sc iv}], C\,{\sc iv}, C\,{\sc iii}] and Mg\,{\sc ii}. Although a 
detailed study of the quasar absorption lines is out of the scope of this paper, several absorption 
features are also evident in the spectra of A and B. In Appendix~\ref{sec:specextevo}, we assess 
the quality of the spectral extraction, as well as the short-term evolution of the quasar spectral 
energy distribution.

\begin{figure}
\centering
\includegraphics[width=9cm]{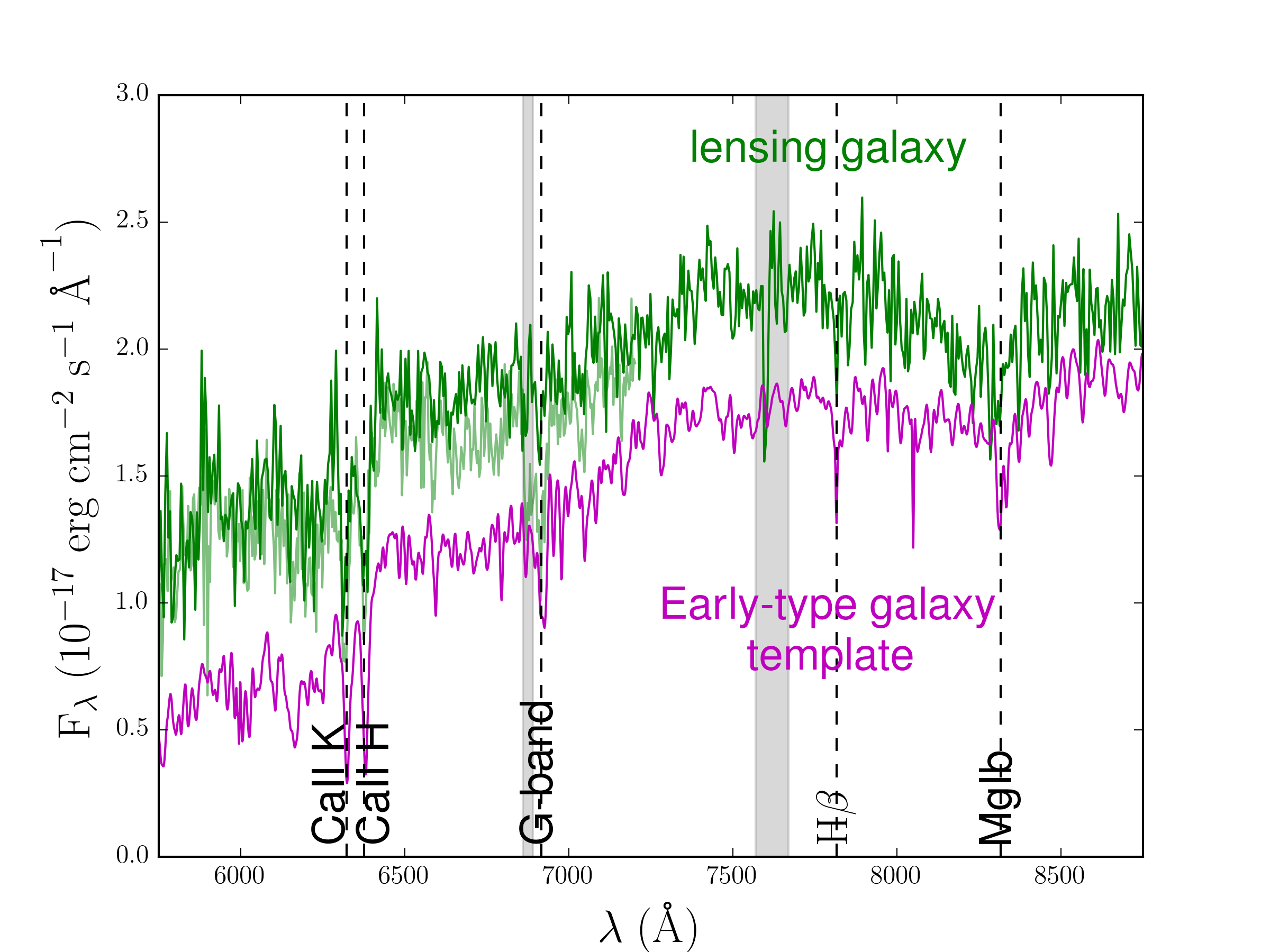}
\caption{GTC-OSIRIS spectra of SDSS J1339+1310G in 2014 and the early-type galaxy template from the 
SDSS database. The fluxes from the R500R and R500B grism data are shown in bright and light green, 
respectively. We note that the spectral edges are particularly noisy and not included here. The 
red-shifted ($z$ = 0.607) spectral template is drawn with magenta lines. Vertical dashed lines are 
associated with absorption features, while grey highlighted regions are related to atmospheric 
artefacts.}
\label{fig:specG}
\end{figure}

Regarding the lensing galaxy, we analysed the new spectra of G in Fig.~\ref{fig:specABG} to better 
understand its nature. The RVSAO/XCSAO IRAF package \citep{kurtz98} was used to perform the
cross-correlation between these spectra and two different early-type galaxy templates. First, the 
improved spectrum from the red grism and the elliptical galaxy template in the Kinney-Calzetti 
Spectral Atlas of Galaxies \citep{kinney96} led to a signal-to-noise ratio in the cross-correlation 
peak \citep[$r_{\rm{cc}}$;][]{tonry79} of about 5, while $r_{\rm{cc}}$ reached a value of $\sim$ 8 
using the spectrum from the blue grism. Second, taking the early-type galaxy template in the SDSS 
database\footnote{\url{http://classic.sdss.org/dr7/algorithms/spectemplates/spDR2-023.gif}}, we 
obtained $r_{\rm{cc}}$ values in the interval $6-7$. The four measurements of the lens redshift (one 
for each combination spectrum-template) ranged between 0.607 and 0.608. In Fig.~\ref{fig:specG}, we 
compare the two galaxy spectra with the red-shifted SDSS template (adopting $z_{\rm{l}}$ = 0.607; see 
below). These GTC-OSIRIS spectra clearly show several absorption features (e.g. the Ca\,{\sc ii} 
$HK$ doublet and the G-band are unambiguously detected with both grisms), which allowed us to 
estimate $z_{\rm{l}}$ values close to 0.606. Thus, the data of G in 2014 lead to $z_{\rm{l}}$ = 0.607 
$\pm$ 0.001 (1$\sigma$ interval). Although the new typical value of the lens redshift differs only in 
a 0.3\% from the previous one in \citetalias{shalyapin14}, we adopt the $z_{\rm{l}}$ for the improved 
spectra. We did not try to fit stellar population synthesis models to the spectra, but the lack of 
[O\,{\sc ii}] emission at 3727 \AA\ (it would be observed at $\sim$ 6000 \AA) is likely indicating a 
lack of star formation in the lensing galaxy \citep[e.g.][]{kennicutt98}.      

\section{Microlensing in the broad line and nuclear continuum emitting regions}
\label{sec:microspec}

GTC-OSIRIS-R500R spectroscopy of the doubly imaged quasar \object{SDSS J1339+1310} in 2013, revealed 
the presence of strong chromatic microlensing of its continuum, which is primarily generated in a 
nuclear region. In \citetalias{shalyapin14}, we used the flux ratio $B/A$ for the cores of three 
emission lines (C\,{\sc iv}, C\,{\sc iii}] and Mg\,{\sc ii}) to estimate the macrolens magnification 
and dust extinction ratios, $M_{\rm{BA}}$ and $\epsilon_{\rm{BA}}$, and thus find the microlensing 
magnification ratio $\mu_{\rm{BA}}$ of light basically coming from the nuclear continuum emitting 
region (NCER). Our main hypothesis in this first paper was that the cores of emission lines are 
dominated by photons arising from the narrow line emitting region (NLER) and the outer parts of the 
broad line emitting region (BLER). Therefore, the line cores would be produced in extended regions 
that are unaffected by microlensing \citep[e.g.][and references therein]{motta12}. 

The GTC-OSIRIS low-noise spectra of the lensed quasar in 2014 cover the whole range of optical 
wavelengths and contain five prominent emission features, that is the three previous lines and two 
additional features at bluer wavelengths (Ly$\alpha$ and Si\,{\sc iv}/O\,{\sc iv}]; see 
Fig.~\ref{fig:specABG}). These new data offer a much richer information than those in 
\citetalias{shalyapin14}, and are useful to check the hypothesis about microlensing-free fluxes of 
line cores (see Sec.~\ref{sec:miclinec} and Sec.~\ref{sec:micline}), as well as to improve our 
knowledge on the extinction and magnification in \object{SDSS J1339+1310} (see Sec.~\ref{sec:micline} 
and Sec.~\ref{sec:micont}). We considered a number of details to analyse the optical spectra of A and 
B, including an accurate estimation of the total and nuclear power-law continua. The spectral energy 
distributions of the A image at 4850$-$7200 \AA\ for the two spectroscopy epochs in 2014 are 
remarkably similar (see Fig.~\ref{fig:specABG} and Appendix~\ref{sec:specextevo}), so we mainly used 
the blue grism data to estimate delay-corrected flux ratios over a very broad wavelength range 
(3600$-$7200 \AA; it was implicitly assumed that the observed constancy of the spectrum of A 
in the $g$ and $r$ bands can be extended to the $u$ band). These delay-corrected ratios were 
complemented with two measurements from the red grism data: flux ratios of the C\,{\sc iii}] and 
Mg\,{\sc ii} line cores (a justification for the use of the two additional single-epoch ratios is 
provided at the end of Appendix~\ref{sec:specextevo}).

\subsection{Microlensing-free fluxes of line cores?}
\label{sec:miclinec}

\begin{figure}
\centering
\includegraphics[width=9cm]{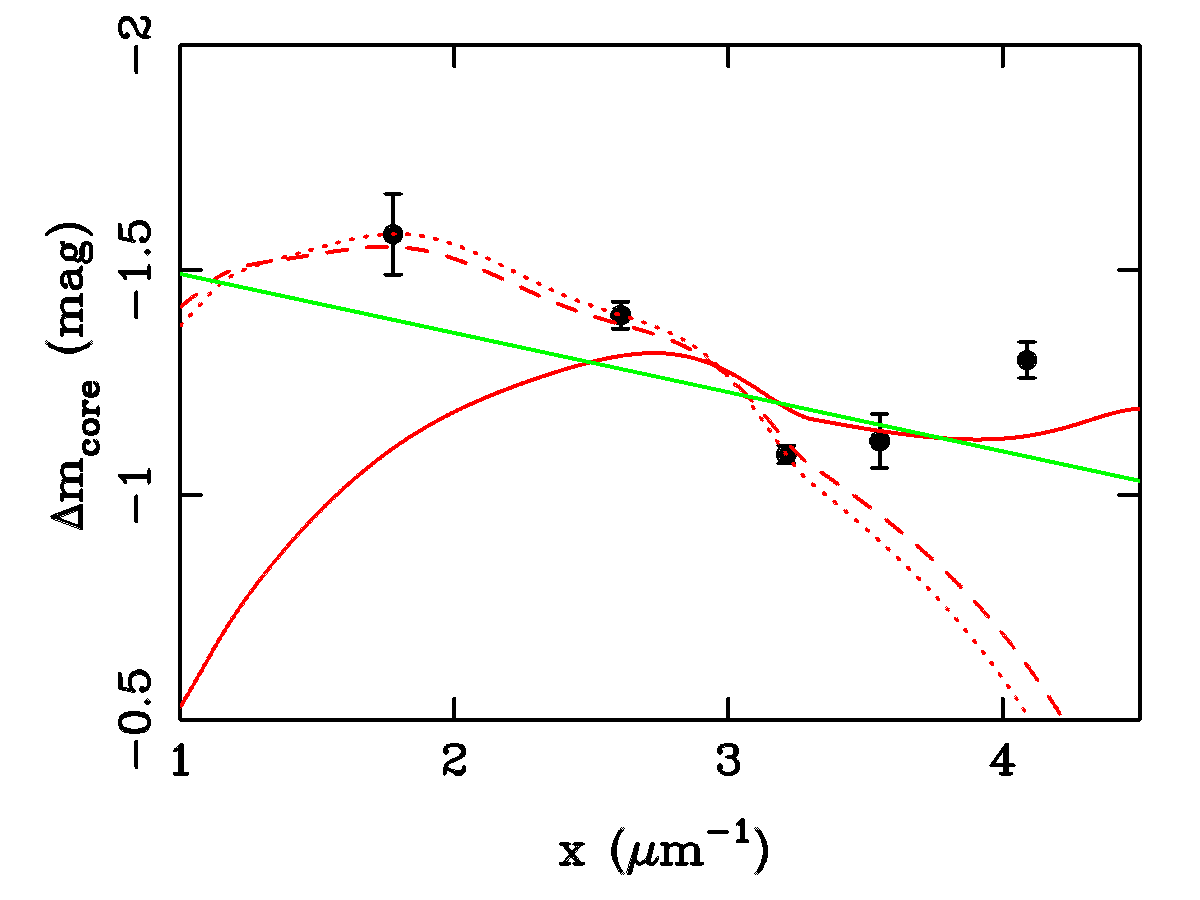}
\caption{Line-core magnitude differences for SDSS J1339+1310. Filled circles with error bars describe
our measurements in the fifth column of Table~\ref{tab:lcfrat}. Instead of $\lambda$ values, the 
x-axis displays the wavelength-dependent quantity $x \propto 1/\lambda$ (see main text). Assuming 
microlensing-free fluxes of line cores and Galactic extinction in the lensing galaxy, we also show 
the best fits using Mg\,{\sc ii} + C\,{\sc iii}] + C\,{\sc iv} data (dotted red line), Mg\,{\sc ii} + 
C\,{\sc iii}] + C\,{\sc iv} + Si\,{\sc iv}/O\,{\sc iv}] data (dashed red line) and the five 
measurements (solid red line). For a standard linear extinction in the lensing galaxy, the solid 
green line traces the best fit to the five magnitude differences.}
\label{fig:magext}
\end{figure}

We initially assumed that the line cores are unaffected by microlensing and only macrolens-extinction 
effects are playing a role \citep[e.g.][and references therein]{motta12}. In this scenario, due to 
differential dust extinction in the lensing galaxy, the line-core flux ratios and the associated 
magnitude differences are expected to have a certain chromatic behaviour. Using a standard formalism 
for the differential extinction of lensed images \citep[e.g.][]{falco99,wucknitz03,elias06}, we 
presented two chromaticity laws in equations 2$-$3 of \citetalias{shalyapin14}, which can be compared 
with the observed magnitude differences in Table~\ref{tab:lcfrat}. Although we were able to perfectly 
fit the Galactic law to the new data for the cores of the Mg\,{\sc ii}, C\,{\sc iii}] and C\,{\sc iv} 
emissions ($\chi^2 \sim$ 0; see the dotted red line in Fig.~\ref{fig:magext}), a large value of 
$\chi^2$ was obtained by adding the magnitude difference for Si\,{\sc iv}/O\,{\sc iv}] ($\chi^2 \sim$ 
10 with 1 degree of freedom; see the dashed red line in Fig.~\ref{fig:magext}). The fitting result 
was even worse when we used the five available differences ($\chi^2 \sim$ 80 with 2 degrees of 
freedom; see the solid red line in Fig.~\ref{fig:magext}).

Apart from the Galactic extinction \citep{cardelli89}, we also considered the simple linear law: $a + 
bx$, $x = (1 + z_{\rm{l}})/\lambda$, as a second variant. This is a reasonable approach for dust 
extinction at $x \sim$ 2$-$4 \citep[e.g.][]{prevot84}, that is in the spectral region of interest (see 
Fig.~\ref{fig:magext}). The standard linear law did not fit significantly better than the Galactic 
law, since the five magnitude differences in Table~\ref{tab:lcfrat} produced a still unacceptable 
chi-square value ($\sim$ 80 with 3 degrees of freedom; see the solid green line in 
Fig.~\ref{fig:magext}). Therefore, our analysis indicated the existence of a conflict between a 
macrolens-extinction scenario and the full dataset of $\Delta m_{\rm{core}}$ values, and thus 
revealed that microlensing is affecting to some extent the emission line cores. Although we assumed 
that the line central features are dominated by photons arising from the NLER and the outer parts of 
the BLER, some of them contain an appreciable number of photons coming from clouds at inner zones of 
the BLER and moving almost perpendicularly to the line of sight (small projected motions). 

\begin{figure}
\centering
\includegraphics[width=9.3cm]{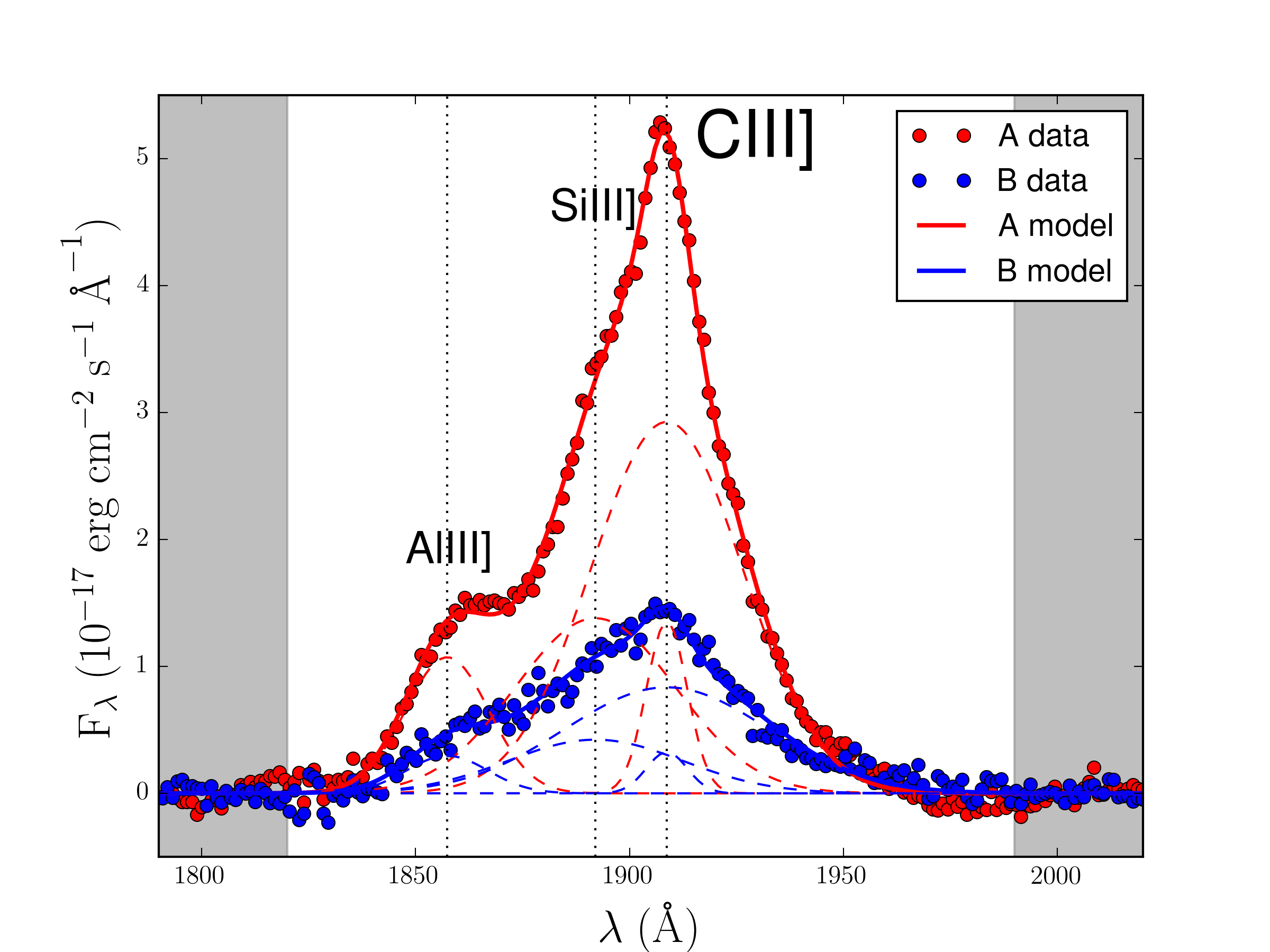}
\includegraphics[width=9.3cm]{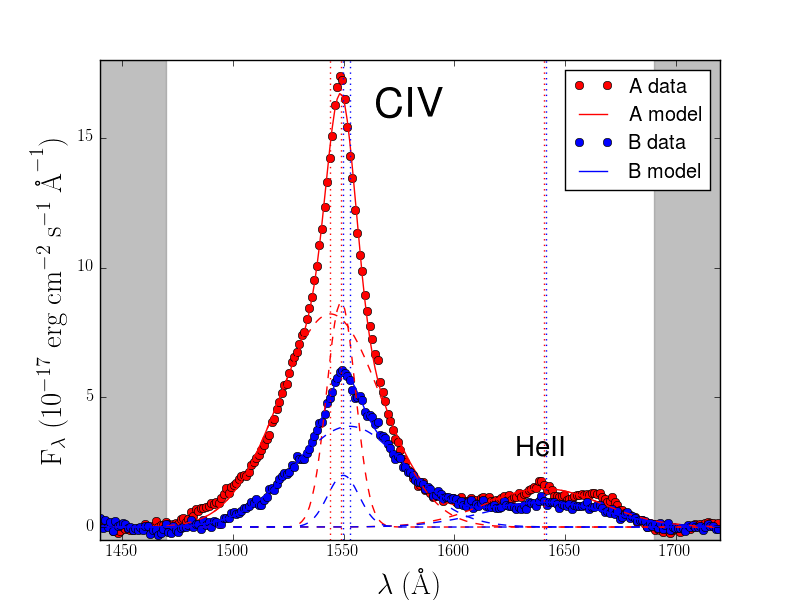}
\includegraphics[width=9cm]{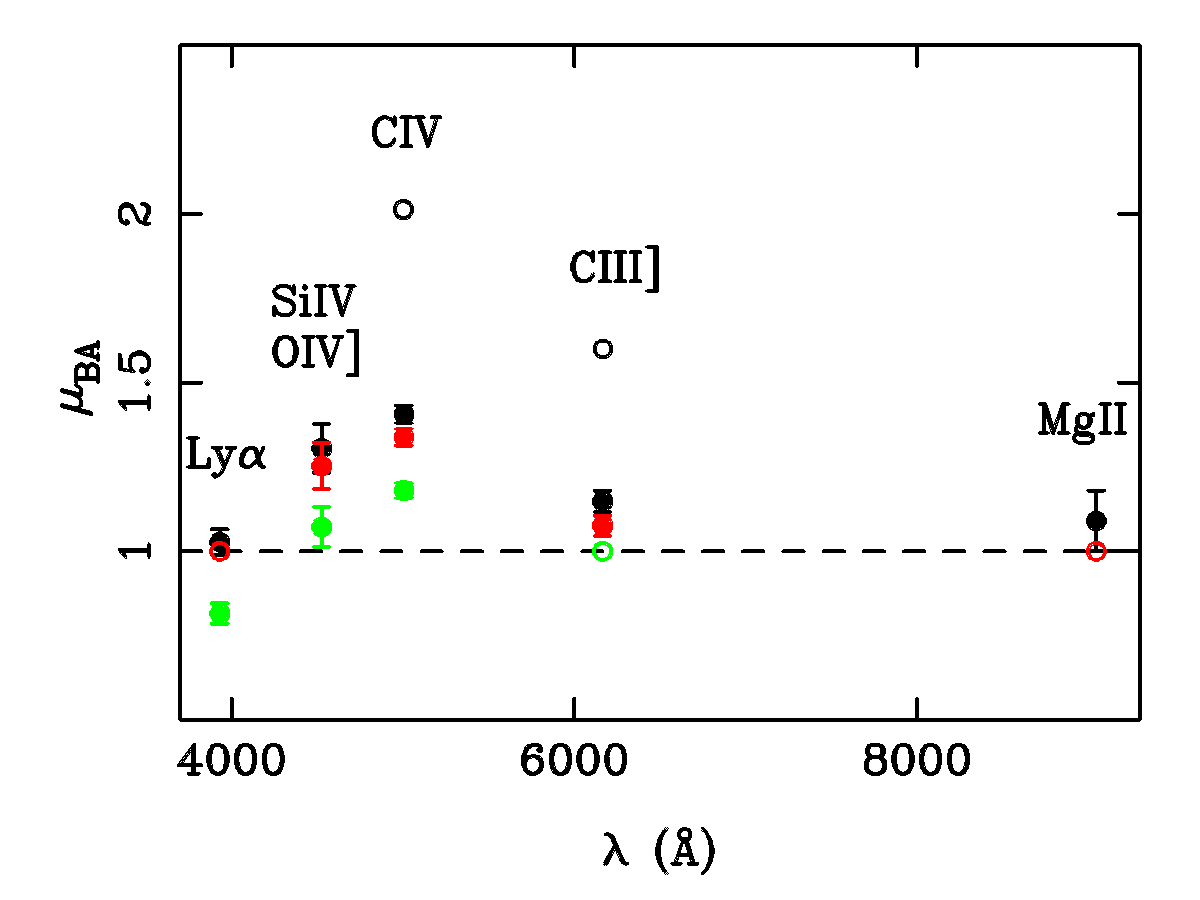}
\caption{Multi-component decomposition of carbon line profiles and microlensing of emission lines. 
The top and middle panels display decompositions into four (C\,{\sc iii}] narrow + C\,{\sc iii}] 
broad + Si\,{\sc iii}] + Al\,{\sc iii}) and three (C\,{\sc iv} narrow + C\,{\sc iv} broad + He\,{\sc 
ii} complex) Gaussians, respectively. The bottom panel incorporates the microlensing magnification 
ratios of the line cores: decomposition of carbon lines (filled black circles; see 
Sec.~\ref{sec:micline}), standard stratification of the BLER (green circles; see 
Sec.~\ref{sec:miclinec}) and non-standard stratification of the BLER (red circles; see 
Sec.~\ref{sec:miclinec}). 
This last panel also shows the microlensing of the C\,{\sc iii}] and C\,{\sc iv} broad emission lines 
(open black circles).}   
\label{fig:micline}
\end{figure}

The BLER is likely stratified, so different atoms or ions are located at different distances from the 
NCER. For example, there is evidence for a correlation between degree of ionization and distance, 
with the high-ionization species much closer to the NCER than those of low-ionization 
\citep[e.g.][]{krolik91,gaskell07}. In this standard stratification of the BLER (SSB), the C\,{\sc 
iv}, Si\,{\sc iv}/O\,{\sc iv}] and Ly$\alpha$ cores could suffer slight microlensing effects, while 
the Mg\,{\sc ii} and C\,{\sc iii}] cores would probably be free of microlensing (see 
Table~\ref{tab:lcfrat}). Hence, the low-ionization, outer lines of Mg\,{\sc ii} and C\,{\sc iii}] 
were used to obtain a macrolens-extinction solution (standard linear law), and remove macrolens and 
extinction effects in the cores of the C\,{\sc iv}, Si\,{\sc iv}/O\,{\sc iv}] and Ly$\alpha$ 
emissions. The best-fit solution was $M_{\rm{BA}}$ = 0.16 and $\epsilon_{\rm{BA}}(V)$ = 1.44, and the 
microlensing magnification ratios of the C\,{\sc iv}, Si\,{\sc iv}/O\,{\sc iv}] and Ly$\alpha$ line 
cores are shown in the bottom panel of Fig.~\ref{fig:micline} (filled green circles). As expected, 
all the $\mu_{\rm{BA}}(\rm{core})$ values are close to one. 

In addition to a SSB, we also considered a non-standard stratification of the BLER (NSB): the typical
distance from the central engine to a specie depends on the total energy that is required to produce 
each excited or ionised atom of such specie (the higher the energy of excitation or ionization, the 
distance is smaller). This naive scenario relies on the idea that the available amount of energy for 
excitation or ionization decreases with distance, leading to outer lines of Mg\,{\sc ii} and Ly$\alpha$ 
(see Table~\ref{tab:lcfrat}). We used these two lines to infer a new macrolens-extinction solution, 
$M_{\rm{BA}}$ = 0.19 and $\epsilon_{\rm{BA}}(V)$ = 1.22, which allowed us to discuss the microlensing 
magnification ratios of the C\,{\sc iii}], C\,{\sc iv} and Si\,{\sc iv}/O\,{\sc iv}] line cores. It 
is noteworthy that we assumed $\mu_{\rm{BA}}(\rm{core})$ = 1 (absence of microlensing) for the 
Mg\,{\sc ii} and Ly$\alpha$ lines (excitation energy of hydrogen and ionization energy of magnesium 
of $\sim$ 10 eV; see the open red 
circles in the bottom panel of Fig.~\ref{fig:micline}), and then obtained a microlensing signal 
$\mu_{\rm{BA}}(\rm{core}) >$ 1 for the other lines (see the filled red circles in the bottom panel of 
Fig.~\ref{fig:micline}), with a clear correlation between the $\delta = \mu_{\rm{BA}}(\rm{core}) - 1$ 
values and the ionization energies (i.e. $\delta/E_{\rm{ion}} \sim$ 0.08/35 (C\,{\sc iii}]), 0.34/80 
(C\,{\sc iv}) and 0.25/60 (Si\,{\sc iv}/O\,{\sc iv}]) eV$^{-1}$). This correlation is fully 
consistent with the NSB scenario.

\subsection{Microlensing in the BLER}
\label{sec:micline}

To estimate a reliable macrolens-extinction solution, the ideal procedure is to use pure narrow lines 
arising from the NLER \citep[e.g.][]{mousta03}, instead of line cores presumably unaffected by 
microlensing (e.g. the SSB and NSB scenarios give two solutions slightly different from each other; 
see above). Unfortunately, these pure narrow lines are not available yet, so we performed the 
decomposition of the carbon lines to get a robust solution. The C\,{\sc iii}] and C\,{\sc iv} 
emission lines are located within the central region of the full range of wavelengths, and their 
signal strengths are enough as to do decompositions into broad and narrow components. With a little 
more detail, our decomposition is based on the blue grism spectra of the lensed quasar \object{SDSS 
J1339+1310}. After de-redshifting the spectra to their rest frame, apart from continuum subtractions, 
each line profile was modelled as a sum of Gaussian contributions (see below). A multi-component 
decomposition of line profiles has been used in many previous works with different aims 
\citep[e.g.][]{wills85,kurasz02,dietrich03,sluse07,marziani10}, and recently, \citet{sluse11} 
identified and studied different components of the C\,{\sc iii}] and C\,{\sc iv} lines in spectra of 
the gravitationally lensed quasar Q2237+0305.

To reproduce the C\,{\sc iii}]$\lambda$1909 profile in each quasar image, after some probes, we 
used two C\,{\sc iii}] components (broad and narrow), plus two blue-wing excesses due to Si\,{\sc 
iii}]$\lambda$1892 and Al\,{\sc iii} $\lambda$1857 emissions \citep[e.g.][]{brother94}. Each of 
these four Gaussian components is characterised by three parameters: central wavelength 
$\lambda_{\rm{c}}$, width (standard deviation) and amplitude. The C\,{\sc iii}], Si\,{\sc iii}] and 
Al\,{\sc iii} emissions were fixed at $\lambda_{\rm{c}}$ = 1908.7, 1892.0 and 1857.4 \AA, 
respectively. We then fitted the model to the A image data. To fit the B image data, the width of the 
C\,{\sc iii}] narrow component was set to be equal to that of the C\,{\sc iii}] narrow component in 
A (i.e. 4.8 \AA), while the rest of parameters (seven) were allowed to vary. Our 4-Gaussian fits: 
C\,{\sc iii}] narrow + C\,{\sc iii}] broad + Si\,{\sc iii}] + Al\,{\sc iii}, are shown in the top 
panel of Fig.~\ref{fig:micline}. These fits yield $(B/A)_{\rm{narrow}}$ = 0.239 and 
$(B/A)_{\rm{broad}}$ = 0.384. 

Regarding the C\,{\sc iv} $\lambda$1549 line profiles, we used 3-Gaussian fits: C\,{\sc iv} narrow + 
C\,{\sc iv} broad + He\,{\sc ii} complex. The so-called He\,{\sc ii} complex is a blend of several 
lines that is described as a single Gaussian centred at $\lambda_{\rm{c}}$ = 1640.7 \AA\ 
\citep[e.g.][]{croom02}. Although we probed more sophisticated descriptions of this complex 
\citep[e.g.][]{fine10}, the best-fit parameters for the C\,{\sc iv} emission did not change in a 
significant way. Thus, we think the single-Gaussian model is enough to account for contamination. The 
fits with three components and eight free parameters are depicted in the middle panel of 
Fig.~\ref{fig:micline}, and resulted in the flux ratios $(B/A)_{\rm{narrow}}$ = 0.260 and 
$(B/A)_{\rm{broad}}$ = 0.523. The C\,{\sc iv} narrow components in the two images have very 
similar position-structure parameters, since their central wavelengths only deviate from each other 
by $\sim$ 1 \AA (i.e. $\lambda_{\rm{c}}(\rm{A})$ = 1548.9 \AA\ and $\lambda_{\rm{c}}(\rm{B})$ = 
1550.0 \AA), and their widths differ in $<$ 1 \AA\ (6.5 \AA\ for A and 7.3 \AA\ for B). Therefore, the 
narrow component in B does not seem to be noticeably affected by the line profile deformation that we 
discuss here below. We also note that there is a significant difference of 8.5 \AA\ between the 
central wavelengths of the two broad components: $\lambda_{\rm{c}}(\rm{A})$ = 1544.0 \AA\ and 
$\lambda_{\rm{c}}(\rm{B})$ = 1552.5 \AA.

Our multi-component decomposition gives two narrow-line flux ratios that can be used to determine a 
reliable macrolens-extinction solution. This solution was $M_{\rm{BA}}$ = 0.17 and 
$\epsilon_{\rm{BA}}(V)$ = 1.28, amid those from the SSB and NSB approaches in 
Sec.~\ref{sec:miclinec}. As $\epsilon_{\rm{BA}}(V)$ is the dust extinction ratio in the $V$ 
band (i.e. $\epsilon_{\rm{B}}(V)/\epsilon_{\rm{A}}(V)$), we find that A is the most reddened image.
After removing macrolens and extinction effects in the broad-line and 
line-core flux ratios, in the bottom panel of Fig.~\ref{fig:micline}, we show new microlensing 
signals $\mu_{\rm{BA}}(\rm{broad})$ (open black circles) and $\mu_{\rm{BA}}(\rm{core})$ (filled black 
circles). The C\,{\sc iv} and Si\,{\sc iv}/O\,{\sc iv}] line cores are clearly affected by 
microlensing, and there is reasonable agreement between the $\mu_{\rm{BA}}(\rm{core})$ values from 
the decomposition of carbon line profiles and the NSB. From the SSB scenario, we inferred biased 
microlensing ratios of the C\,{\sc iv}, Si\,{\sc iv}/O\,{\sc iv}] and Ly$\alpha$ line cores (filled 
green circles in the bottom panel of Fig.~\ref{fig:micline}). This bias is due to the wrong 
assumption $\mu_{\rm{BA}}(\rm{core})$ = 1 for the C\,{\sc iii}] line (open green circle), which is 
not supported by the decomposition in the top and middle panels of Fig.~\ref{fig:micline}. The 
C\,{\sc iii}] BLER is not so extended as assumed in our framework for the SSB, and 
$\mu_{\rm{BA}}(\rm{broad})$ = 1.6 for this line. We also detected an important microlensing effect on 
the C\,{\sc iv} BLER, since B is amplified (relative to A) by a factor of two.  

\begin{figure}
\centering
\includegraphics[width=9cm]{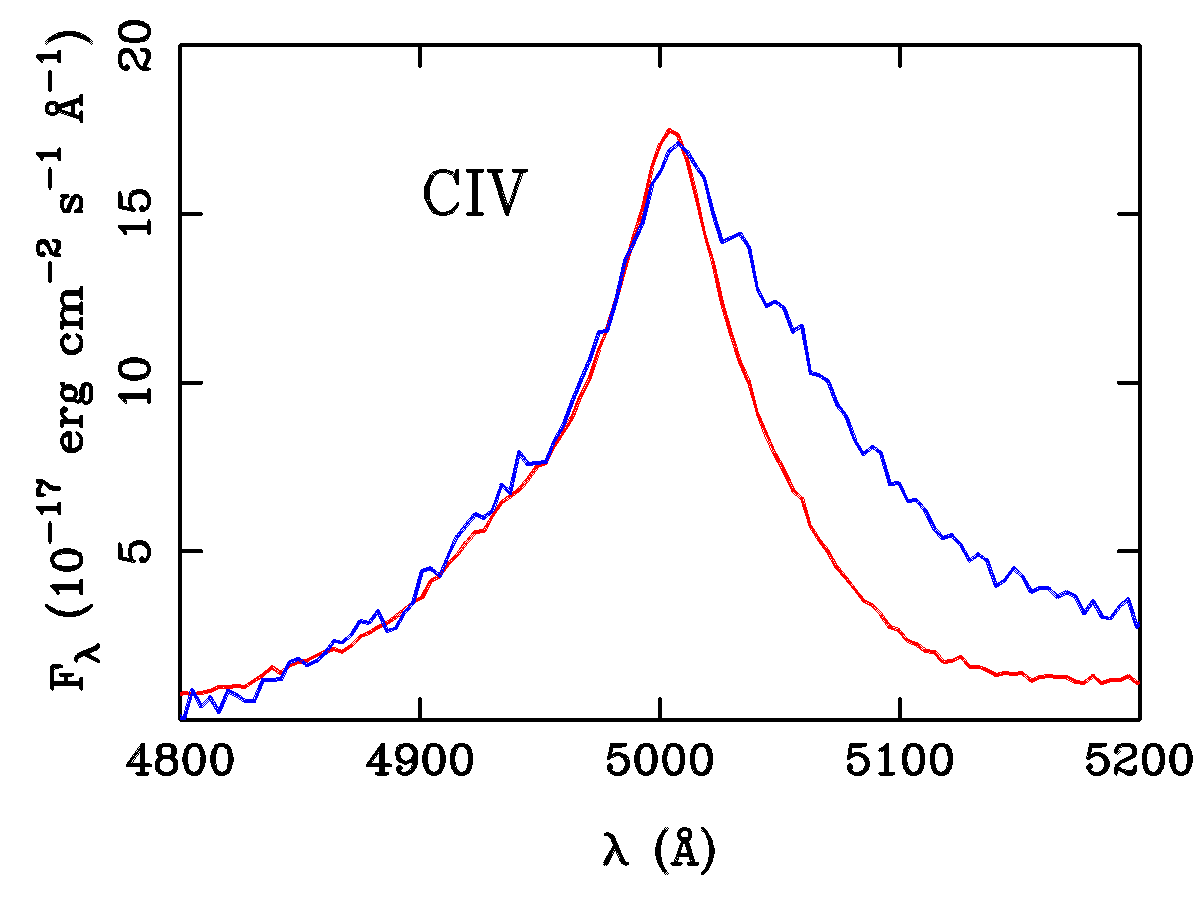}
\caption{C\,{\sc iv} emission line profiles for SDSS J1339+1310. The red line represents the profile 
for A, and the blue line traces the profile for B multiplied by a factor to match the line core of A.  
A power-law continuum has been subtracted from the spectra of A and B \citep[e.g.][]{richards04}.}
\label{fig:profCIV}
\end{figure}

In general, the BLER of lensed quasars suffers weak microlensing effects 
\citep[e.g.][]{motta12,guerras13}. However, the C\,{\sc iv} BLER of \object{SDSS J1339+1310} is 
notably microlensed, which may produce a C\,{\sc iv} line profile for B different to that for A 
\citep[e.g.][]{nemiroff88,schneider90,abajas02,lewis04}. This microlensing-induced distortion of the 
line shape was observed in the quadruple quasar SDSS J1004+4112 at several epochs: when comparing the 
C\,{\sc iv} emission line in its images A and B, the A image shows blue wing enhancements and red 
wing diminutions \citep[relative to B; e.g.][]{richards04,motta12}. Here, we also checked whether 
this type of distortion occurs or not in \object{SDSS J1339+1310}. In Fig.~\ref{fig:profCIV}, the red 
line traces the profile for A, and the blue line describes the profile for B multiplied by the 
inverse of the line-core flux ratio in Table~\ref{tab:lcfrat}. Despite good agreement between both 
blue wings, there is a strong enhancement in the red wing of B relative to that of A. This result 
indicates that the C\,{\sc iv} BLER of \object{SDSS J1339+1310} does not have a spherically symmetric 
structure \citep[e.g.][and references therein]{sluse12}. 

\begin{figure}
\centering
\includegraphics[width=9cm]{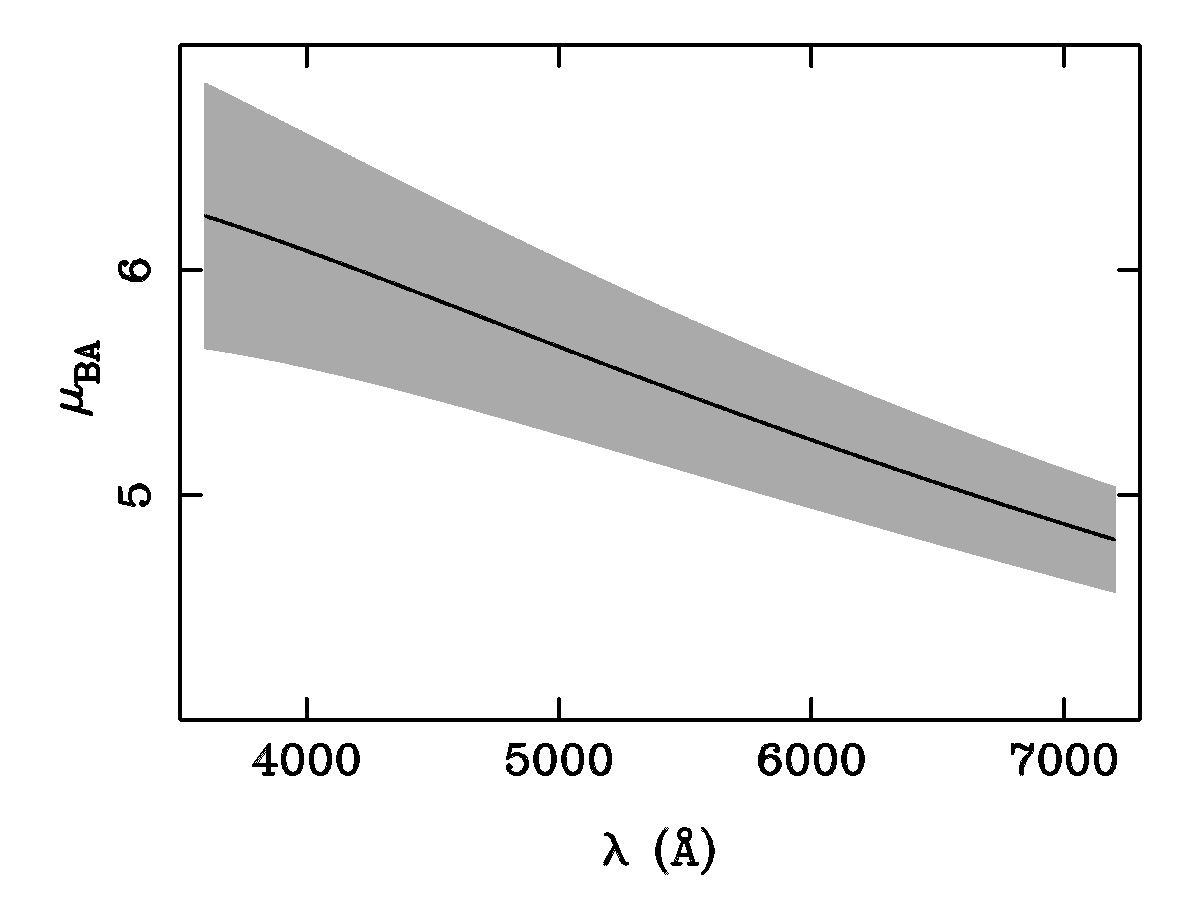}
\caption{Microlensing magnification ratio of the nuclear continuum. We use the best power-law fits to
the continuum of A and B (black line), as well as the 1$\sigma$ bands for these fits (grey region).}
\label{fig:miccont}
\end{figure}

\subsection{Microlensing in the NCER}
\label{sec:micont}

In comparison with the BLER, the NCER should be more strongly affected by microlenses because its 
smaller size \citep[e.g.][]{schneider90,schneider06}. In \citetalias{shalyapin14}, we discussed the 
microlensing magnification ratio of the continuum at $\sim$ 5000$-$9000 \AA, and found that B is 
amplified (relative to A) by a factor of about 3$-$5, with larger amplifications at shorter 
wavelengths. However, in this first analysis, we considered the total continuum from the NCER and the
BLER (Balmer + iron forest), and the time delay between images was not taken into account. Our new
spectra at two epochs separated by $\sim$ 50 d (see Sec.~\ref{sec:spec}), allowed us to properly 
compare the continuum of A and B. In the overlapping spectral region between the blue and red grisms, 
the spectra for A were identical at both epochs. Hence, we have concentrated on the blue grism 
spectra for A and B to estimate the ratio $(B/A)_{\rm{cont}}$. The trailing image B at a given 
observing time $t$ is thus compared to the leading image A at $t - \Delta t$, where $\Delta t \sim$ 
50 d basically coincides with the predicted and measured time delay (see Sec.~\ref{sec:intro} and 
Sec.~\ref{sec:delmicvar}). To calculate $(B/A)_{\rm{cont}}$, we used the nuclear power-law continuum 
instead of the total one. Moreover, the macrolens-extinction effects were corrected through the 
solution in Sec.~\ref{sec:micline}. The final microlensing spectrum $\mu_{\rm{BA}}(\rm{cont})$ is 
shown in Fig.~\ref{fig:miccont}. In this figure, we see the expected chromaticity, with a typical 
magnification ratio of $\sim$ 5.7 at 5000\AA\ (i.e. about three times $\mu_{\rm{BA}}(\rm{broad})$ for 
the C\,{\sc iv} emission line). The microlensing signal $\mu_{\rm{BA}}(\rm{cont})$ in \object{SDSS 
J1339+1310} clearly exceeds those observed in most of lensed quasars for which $\mu_{ij}(\rm{cont}) 
<$ 4 \citep[$i$ = A, B... and $j$ = A, B... with $i \neq j$; e.g.][]{sluse12,rojas14}. 

\section{Light curves in the $r$ band}
\label{sec:lcur}

Immediately after the discovery of the double quasar \object{SDSS J1339+1310} \citep{inada09}, we 
began a robotic monitoring programme with the LT in the SDSS $r$ passband \citep{goico12}. Our 
photometric 
observations were performed in 2009 and 2012$-$2015, and from 13 January to 8 February 2016, so they 
span five seasons of $\sim$ 4$-$6 months each (tipically from January or February to July) and the 
beginning of the current season. For most of the observing nights, we set the exposure time to 600 s. 
Usually this 600s exposure was divided into two 300 s sub-exposures or four 150 s sub-exposures. The 
$r$-band frames in 2009 and 2012 were obtained with the RATCam optical CCD camera, which was 
decommissioned a few years ago. This camera had a $4\farcm6\times4\farcm6$ field of view and a pixel 
scale of $\sim 0\farcs27$ (binning 2$\times$2). All observations in 2013$-$2016 were made using the 
new IO:O camera. IO:O is characterised by a $10\arcmin\times10\arcmin$ field of view and a pixel 
scale of $\sim 0\farcs30$ (binning 2$\times$2). Apart from basic pre-processing tasks included in the 
LT pipelines, we cleaned cosmic rays and interpolated over bad pixels using bad pixel masks. The 
global database consists of 392 individual frames\footnote{All pre-processed frames will be soon 
publicly available on the GLENDAMA archive at \url{http://grupos.unican.es/glendama/database/} 
\citep{goico15}}. 

\begin{figure}
\centering
\includegraphics[width=9cm]{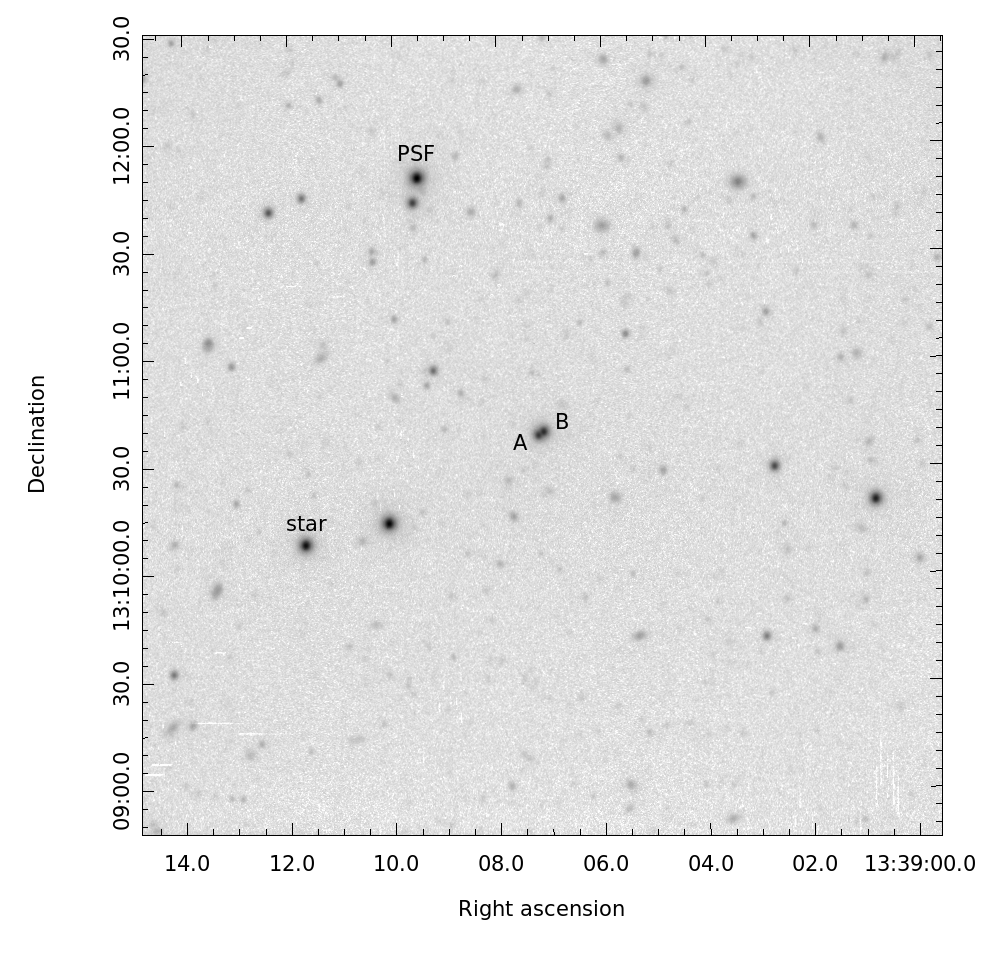}
\caption{$r$-band LT imaging of SDSS J1339+1310. After combining the best RATCam frames in 2009 
(total exposure time = 6900 s), we added labels to the two quasar images (A and B) and the most 
relevant stars in the field of view: PSF star (PSF) and control star (star).}
\label{fig:LTframe}
\end{figure}

At a first stage, a crowded-field photometry pipeline produced relative magnitudes of the quasar 
images in each individual frame. This photometric pipeline is based on IRAF tasks and the IMFITFITS 
software \citep{mcleod98}. We obtained relative fluxes (magnitudes) of A and B through point-spread 
function (PSF) fitting, using the unsaturated star with $r$ = 16.866 mag in the vicinity of the lens 
system as the PSF star\footnote{Other PSF stars led to similar results} (see Fig.~\ref{fig:LTframe}). 
This PSF star is also taken as reference for differential photometry. Our photometric model included 
A and B (two PSFs), the lensing galaxy G (a de Vaucouleurs profile convolved with the PSF) and a 
constant background. We also incorporated constraints in the last column of Table 1 of 
\citetalias{shalyapin14}: the relative positions of B and G (with respect to A), and the structure 
parameters of G. In a first iteration, we applied the code to the frames with the best values of 
seeing and signal-to-noise ratio ($SNR$), allowing the galaxy-to-PSF star ratio ($G/PSF$) to be free. 
We then derived $G/PSF$ = 0.035 (i.e. $r$ = 20.5 mag for G), and adopted this ratio in a second 
iteration. In Fig.~\ref{fig:LTframe}, we show a stacked frame of 
\object{SDSS J1339+1310} with the LT in the $r$ band. This includes, among other objects, the lens 
system (central region), as well as a control star with $r$ = 17.360 mag (down and to the left of the
lens system) and the PSF star. 

\begin{figure}
\centering
\includegraphics[width=9cm]{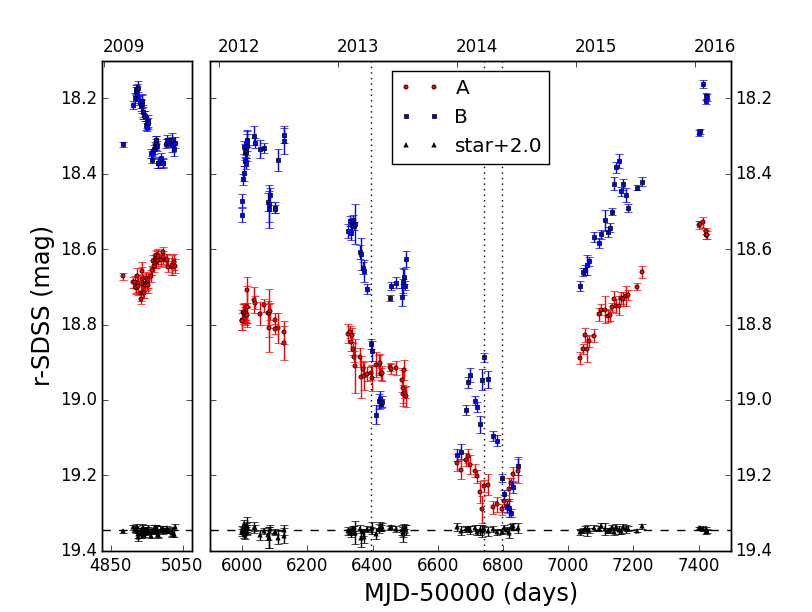}
\caption{LT light curves of both quasar images and the control star. The stellar data are offset by 
+2.0 mag to facilitate comparison. The vertical dotted lines correspond to our spectroscopic 
observations with the GTC in 2013 and 2014.}
\label{fig:lightc}
\end{figure}

In a second stage, we derived final light curves (SDSS magnitudes) for A, B and the control star. We 
have removed 26 individual frames leading to outliers in the initial light curves of the quasar. 
These outlier frames were discarded on basis of objective criteria. First, by visual inspection of 
the brightness records, we found that the magnitudes of A and B strongly deviate from adjacent data.
Deviations of A and B occur in the same direction or in opposite directions (likely due to crosstalk 
between both images), and exceed by more than three times the preliminary typical errors. Second, a 
careful data analysis indicated that either the PSF fitting of the lens system led to a large 
$\chi^2$ value, the image quality was poor or both things hapenned. After this selection of frames, 
we have combined magnitudes measured on the same night. To estimate typical magnitude errors, we used 
the root-mean-square deviations between magnitudes on consecutive nights. The resulting uncertainties 
were 0.019 (A), 0.015 (B) and 0.008 (star) mag. As expected, the typical error in the star agrees 
with its standard deviation of 0.009 mag, and fainter sources have larger uncertainties. We also 
expect, on theoretical grounds, the existence of a correlation between the error $\sigma_m$ and the 
inverse of the $SNR$ \citep[e.g.][]{howell06}. Hence, after calculating the average $SNR$ for the A 
image at the epochs used to determine typical errors, $SNR$(typ), uncertainties on a nightly basis 
were estimated as $\sigma_m = \sigma_m$(typ) $\times$ [$SNR$(typ)/$SNR$]. 

The final light curves of A, B and the star are available in tabular format at the CDS$^3$: Table 3 
includes $r$-SDSS magnitudes $m$ and errors $\sigma_m$ at 143 epochs. Column 1 lists the observing 
date (MJD$-$50\,000), Columns 2 and 3 give $m$ and $\sigma_m$ values for the quasar image A, Columns 
4 and 5 give $m$ and $\sigma_m$ values for the quasar image B, and Columns 6 and 7 indicate the 
magnitudes and their errors for the star. The optical brightness records are also shown in 
Fig.~\ref{fig:lightc}. The quasar exhibits significant time variability, while the horizontal dashed 
line represents the constant flux of the control star. We also remark that the short-timescale signal 
in B is not due to spurious photometry. In the $r$ band, G is $\sim$ 2 mag fainter than B and the sky 
background does not play a dramatic role. Thus, even using a crude photometric model including only 
the two quasar images (two PSFs), we found light curves similar to those in Fig.~\ref{fig:lightc}. 

\section{Time delay and $r$-band microlensing variability}
\label{sec:delmicvar}

As noted in \citetalias{shalyapin14}, our original aim was to determine the time delay of the system 
from a few observing seasons, since the expected delay is less than two months. However, the LT light 
curves over the first seasons did not permit us to measure the delay between A and B. Now, in 
Fig.~\ref{fig:lightc}, we see that A and B appreciably vary in parallel on a long timescale. This 
means that the long-timescale variations are primarily originated in the distant quasar (intrinsic 
origin). In addition, there is also clear evidence of microlensing (extrinsic) variability, for example 
several sharp events in B are not seen in A. Although a time delay measurement in presence of 
microlensing is a relatively complex task, the extrinsic variability can be reasonably described 
either by a suitable type of functions \citep[e.g.][]{tewes13a} or through intensive microlensing 
simulations \citep[e.g.][]{hainline13}. The last method (simulations) was our initial option 
\citepalias[see conclusions of][]{shalyapin14}, but it is computationally expensive, and provides 
results that are not better than those from the first technique (see Sec.~\ref{sec:intro}). 
Therefore, here, we focus on the first approach. This yields a microlensing light curve in the $r$ 
band, which, together with the microlensing spectrum in Fig.~\ref{fig:miccont}, can be compared with 
numerical simulations. Such a comparison is out of the scope of this paper, despite its great 
interest to constrain the $r$-band size and wavelength-dependent structure of the quasar accretion 
disk \citep[e.g.][]{bate08,eigenbrod08,morgan10,mosquera11,motta12}. 

We use two different approaches to determine the time delay of \object{SDSS J1339+1310}, one 
including seasonal microlensing variability (Sec.~\ref{sec:delseamic}) and the other incorporating 
microlensing variations on all timescales (Sec.~\ref{sec:delsplmic}). These approaches and their 
associated methods are useful to track for systematic errors. The final measurement of the delay and 
the $r$-band microlensing variability are discussed in Sec.~\ref{sec:delfinmicvar}. 

\subsection{Time delay: seasonal microlensing}
\label{sec:delseamic}

\begin{figure}
\centering
\includegraphics[width=9cm]{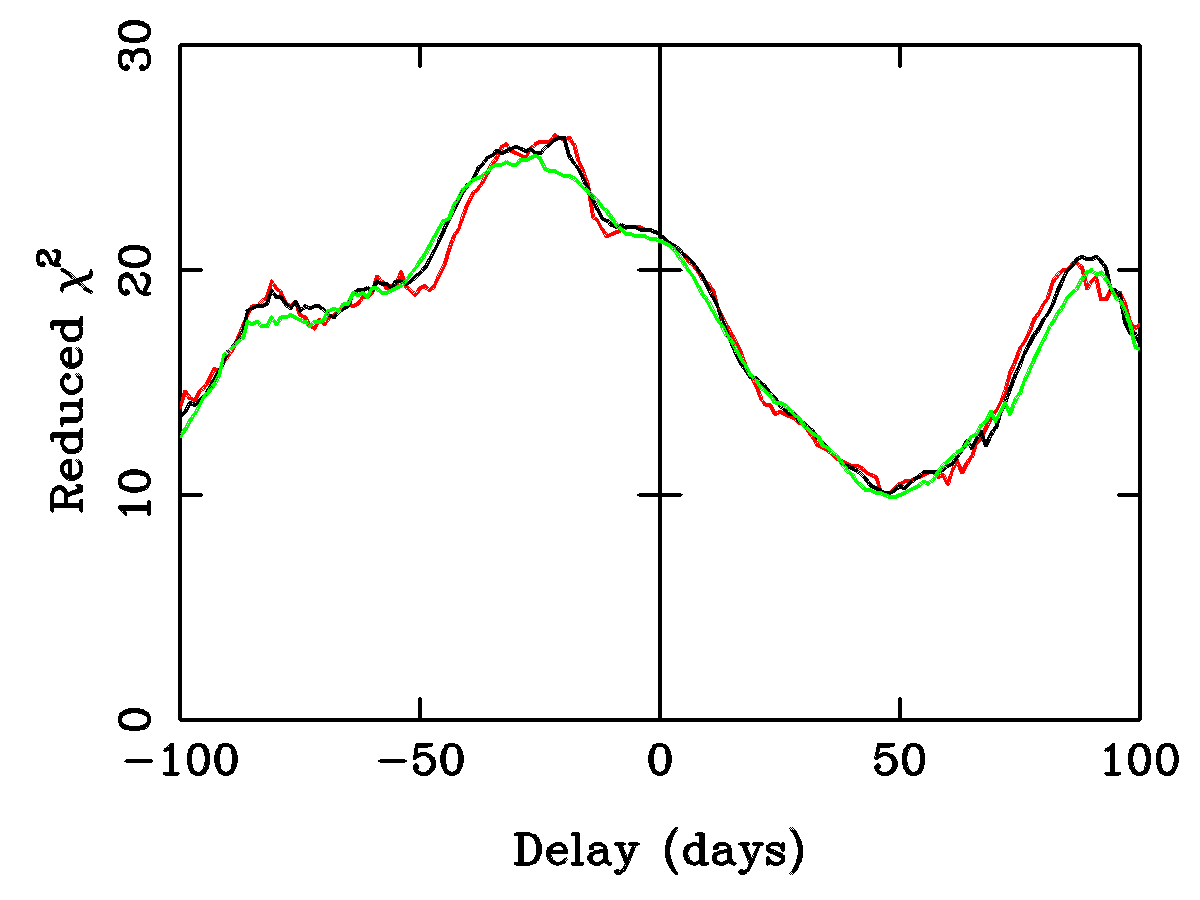}
\caption{$\hat{\chi}^2$-delay relationships for bin semisizes $\alpha$ = 15 (red), 20 (black), and 25 
(green) d.}
\label{fig:chi2_step}
\end{figure}

\begin{figure}
\centering
\includegraphics[width=9cm]{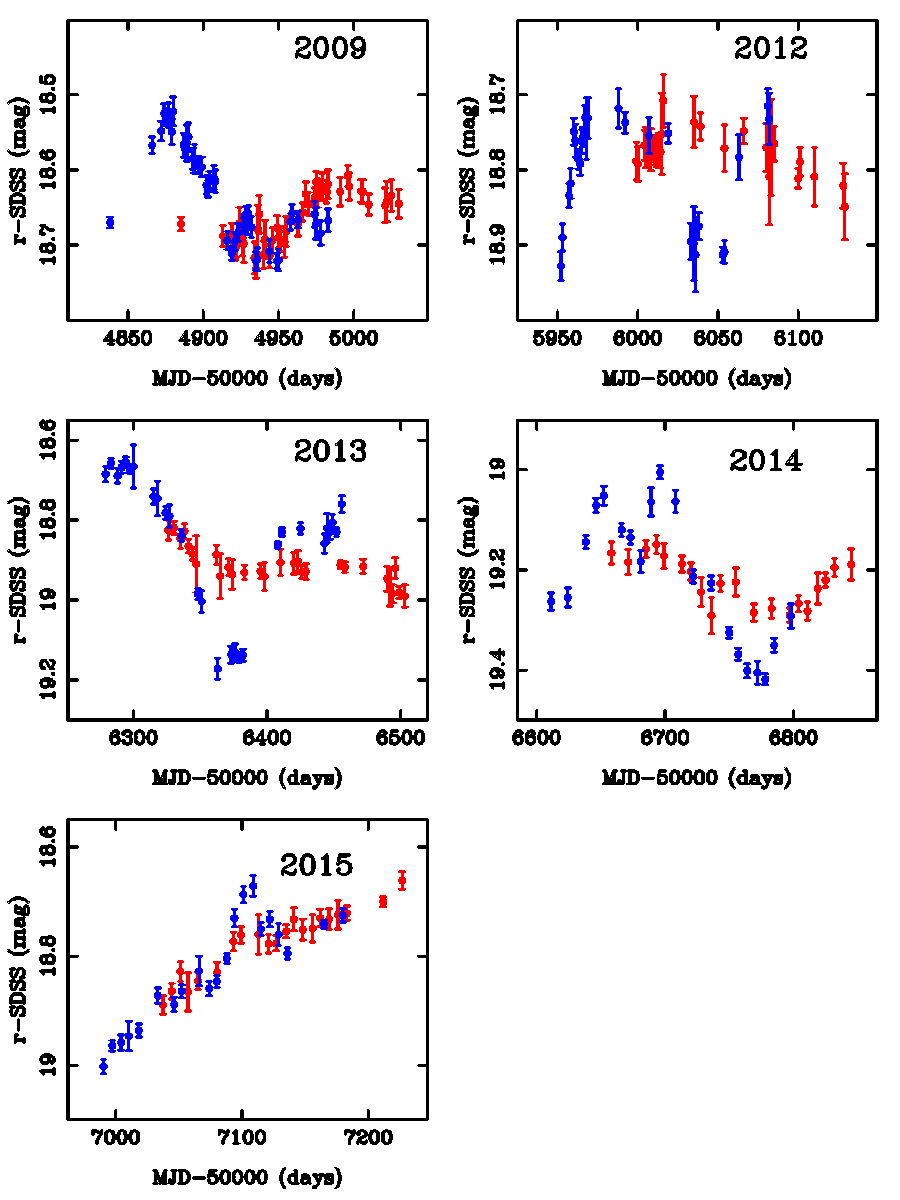}
\caption{Combined light curve in the $r$ band from the $\hat{\chi}^2$ minimization with a step 
function-like microlensing. The A curve (red circles) and the magnitude- and time-shifted B curve 
(blue circles) are drawn together (see main text).}
\label{fig:clc_step}
\end{figure}

We initially performed a reduced chi-square ($\hat{\chi}^2$) minimization to match the light curves 
of A and B over the period 2009$-$2015. The data in early 2016 are only used to demonstrate the 
parallel brightening of both images in recent years. Our $\hat{\chi}^2$ method was based on a 
comparison between the curve A and the time-shifted curve B (shifted forward or backward in 
$\leq$ 100 d), using bins with 
semisize $\alpha$ in B \citep[e.g.][]{ullan06}. To account for the presence of long-timescale 
extrinsic variability, we incorporated six free parameters: a time delay ($\Delta t_{\rm{AB}} = 
\tau_{\rm{B}} - \tau_{\rm{A}}$) and a magnitude offset for each of the five seasons in 2009 and 
2012$-$2015. From the formal point of view, $\Delta m_{\rm{AB}}(t) = m_{\rm{B}}(t + \Delta 
t_{\rm{AB}}) - m_{\rm{A}}(t)$ is assumed to be constant within a given season, but it can vary from 
season to season. After analysing the $\hat{\chi}^2$-delay relationships for different values of 
$\alpha$, we found that 10 $< \alpha \leq$ 30 d lead to reasonably smooth trends with global minima 
at 46$-$49 d (see Fig.~\ref{fig:chi2_step}). Thus, $\alpha$ = 20 d gives a best solution 
$\Delta t_{\rm{AB}}$ = 47 d, $\Delta m_{\rm{AB}}(2009)$ = $-$0.349 mag, $\Delta m_{\rm{AB}}(2012)$ = 
$-$0.419 mag, $\Delta m_{\rm{AB}}(2013)$ = $-$0.133 mag, $\Delta m_{\rm{AB}}(2014)$ = $-$0.118 mag 
and $\Delta m_{\rm{AB}}(2015)$ = $-$0.304 mag ($\hat{\chi}^2 \sim$ 10; see the black line in 
Fig.~\ref{fig:chi2_step}). The combined light curve for this best solution is displayed in the five 
panels of Fig.~\ref{fig:clc_step}. The existence of short-timescale microlensing events in all 
seasons does not allow us to get a best fit with $\hat{\chi}^2 \sim$ 1, since the intra-seasonal 
microlensing works as an additional noise that is not taken into account in the denominator of the 
$\hat{\chi}^2$ addends. 

\begin{figure}
\centering
\includegraphics[width=9cm]{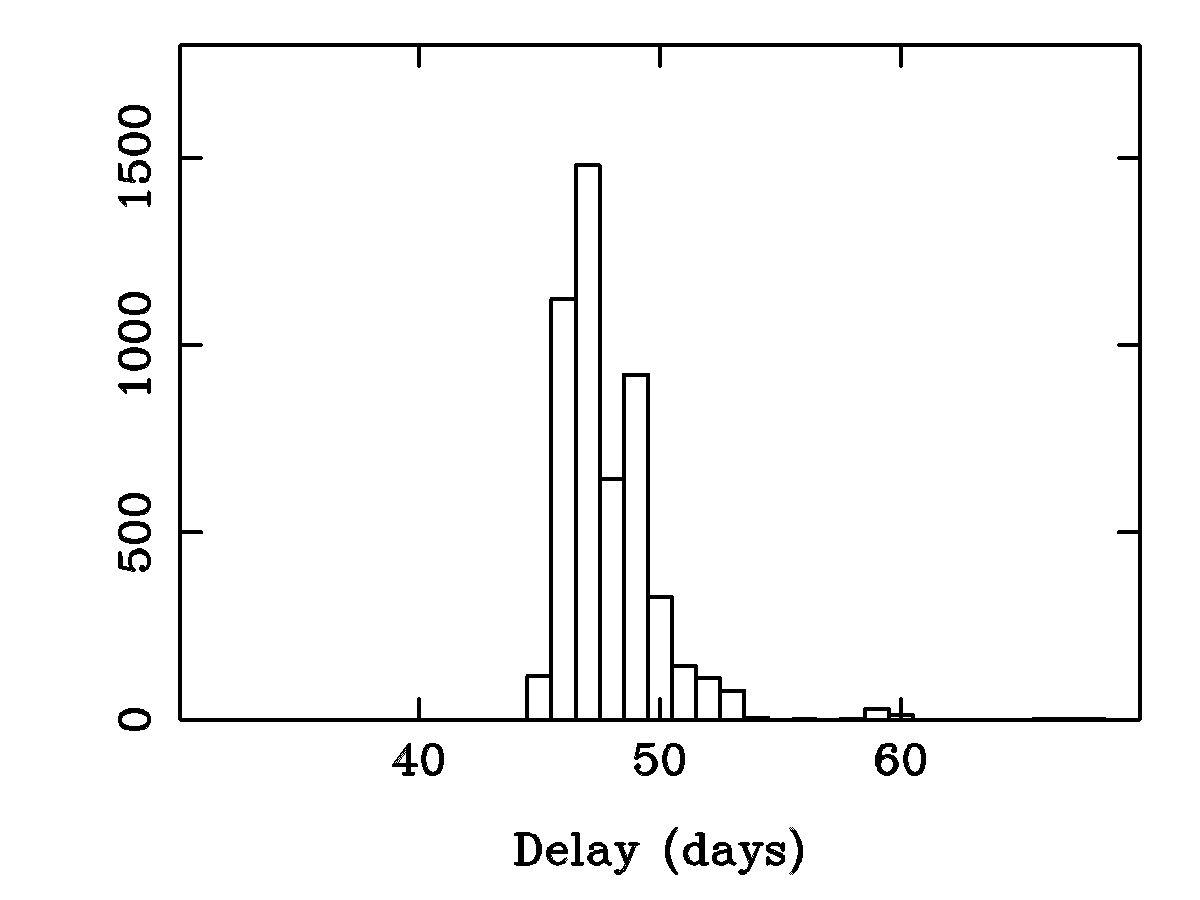}
\includegraphics[width=9cm]{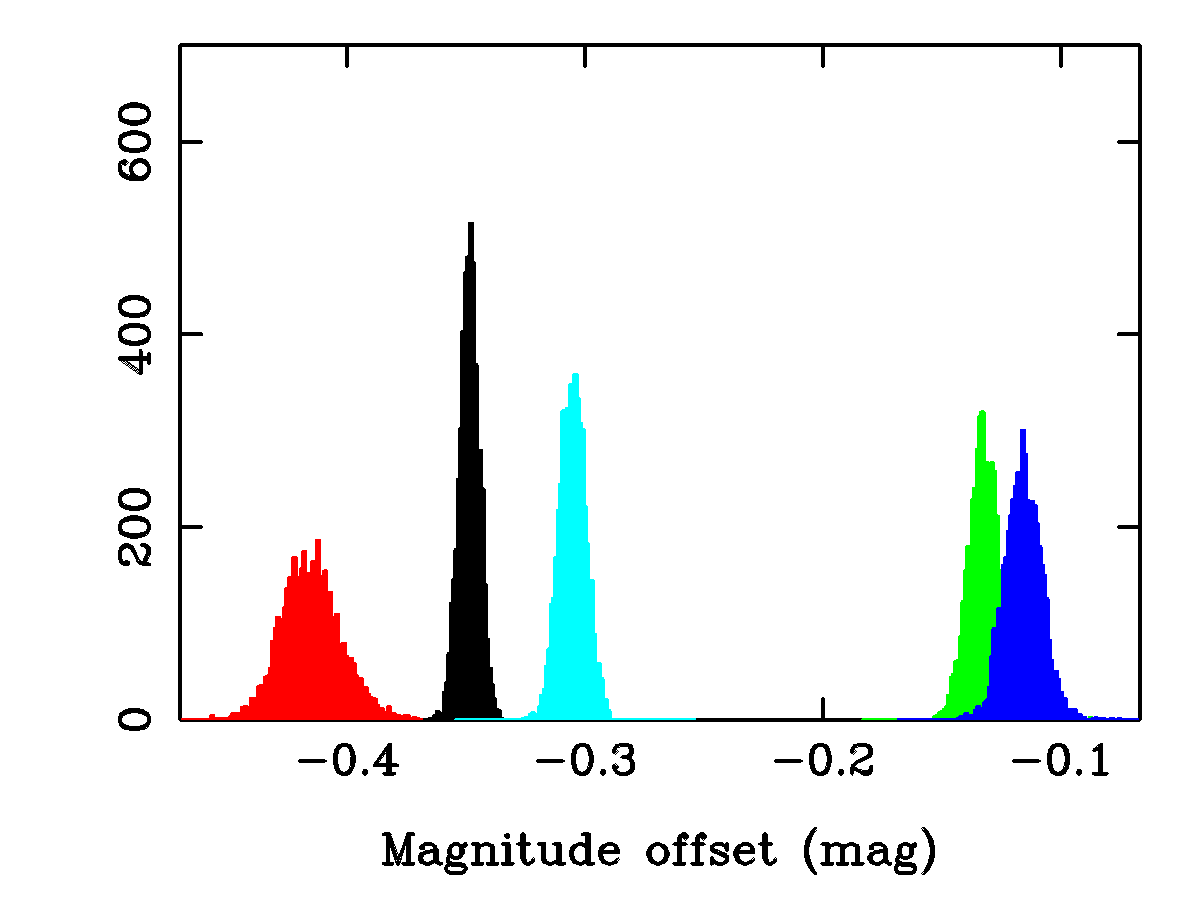}
\caption{Histograms from 1000 pairs of synthetic curves AB and the $\hat{\chi}^2$ minimization 
($\alpha$ = 18$-$22 d) with a step function-like microlensing. The top panel shows the best solutions 
of the time delay. The bottom panel displays the best solutions of the magnitude offsets: 2009 
(black), 2012 (red), 2013 (green), 2014 (blue) and 2015 (cyan).}  
\label{fig:histo_step}
\end{figure}

We used a simple approach to generate synthetic light curves and estimate parameter errors. To obtain 
a pair of synthetic curves AB, the observed magnitudes were modified by additive random quantities 
(i.e. realisations of normal distributions around zero, with standard deviations equal to the 
measured uncertainties). We produced 1000 pairs of synthetic curves, and the $\hat{\chi}^2$ 
minimization ($\alpha$ = 18$-$22 d) with a step function-like microlensing was applied to them. The 
distributions of delays and magnitude offsets are presented in Fig.~\ref{fig:histo_step}, and we pay
special attention to the top panel including the delay histogram. Our 1$\sigma$ measurement of the 
delay is $\Delta t_{\rm{AB}}$ = 47$^{+2}_{-1}$ d, while 45 $\leq \Delta t_{\rm{AB}} \leq$ 51 d is the 
2$\sigma$ (95\%) confidence interval. Although we only consider a 5-d interval of $\alpha$ 
values to account for the intrinsic variance of the technique, this is enough. From the LT light 
curves, we derived distributions of delays for $\alpha$ = 18$-$22 d and for a broader range of values 
of $\alpha$ (12$-$28 d), and found similar standard deviations of $\sim$ 1 d.    

We also used the dispersion ($D^2$) technique \citep{pelt94,pelt96} to check for possible biases 
in the delay estimation from the $\hat{\chi}^2$ method. The $D^2$ minimization is characterised by a 
decorrelation length ($\delta$) and has several variants \citep[mainly the $D^2_{4,1}$ and 
$D^2_{4,2}$ estimators introduced by][]{pelt96}. The key idea of our $\hat{\chi}^2$ method is to take 
the light curve of A as a template for smooth (mainly intrinsic) variability, and compare the A 
magnitudes with binned magnitudes of B. However, the $D^2$ technique does not differenciate between A 
and B, and it works as a kind of average between the $\hat{\chi}^2$ version that we used, and another 
complementary version based on a comparison between the B magnitudes and binned magnitudes of A. From 
the $D^2_{4,1}$ estimator including a step function-like microlensing, we derived minima around an 
average value of 47.5 d ($\delta$ = 25$-$35 d). These independent results confirm a time delay of 
$\sim$ 47 d when only long-timescale extrinsic variability is taken into account. The influence of 
the microlensing model on the delay estimation is discussed in the next subsections.

\subsection{Time delay: spline-like microlensing}
\label{sec:delsplmic}

Although long-timescale microlensing is clearly detected and measured in the bottom panel of 
Fig.~\ref{fig:histo_step}, there are also short-timescale microlensing events that were not modelled 
and might play a role in the determination of the time delay. In Fig.~\ref{fig:clc_step}, we see that
extrinsic magnitude fluctuations on timescales of 50$-$100 d are usual, and they are even faster 
than the intra-seasonal intrinsisc variations. To account for this fast extrinsic variability, we 
used the PyCS software\footnote{PyCS is distributed by the COSMOGRAIL project, and it is available at 
\url{https://github.com/COSMOGRAIL/PyCS}} \citep{tewes13a,bonvin16}. Thus, the intrinsic and 
extrinsic variations were described by free-knot splines. Each cubic spline was parametrised by the 
knot epochs and the associated coefficients, and a $\chi^2$ minimization allowed us to simultaneously 
fit the time delay and the splines. After some tests, we took a knot step $\eta_{\rm{int}}$ = 100 d 
for the intrinsic spline because this step leads to a good fit of the light curve A. For the 
extrinsic spline, $\eta_{\rm{ext}}$ = 100 d gave rise to rough fits to the data: $\chi^2 \sim$ 600 
with $N$ = 2$\times$137 = 274 data points. The time delay through this rough spline-like microlensing 
was $\sim$ 47 d, in very good agreement with our initial measurement using a step function-like 
microlensing. 

\begin{figure}
\centering
\includegraphics[width=9cm]{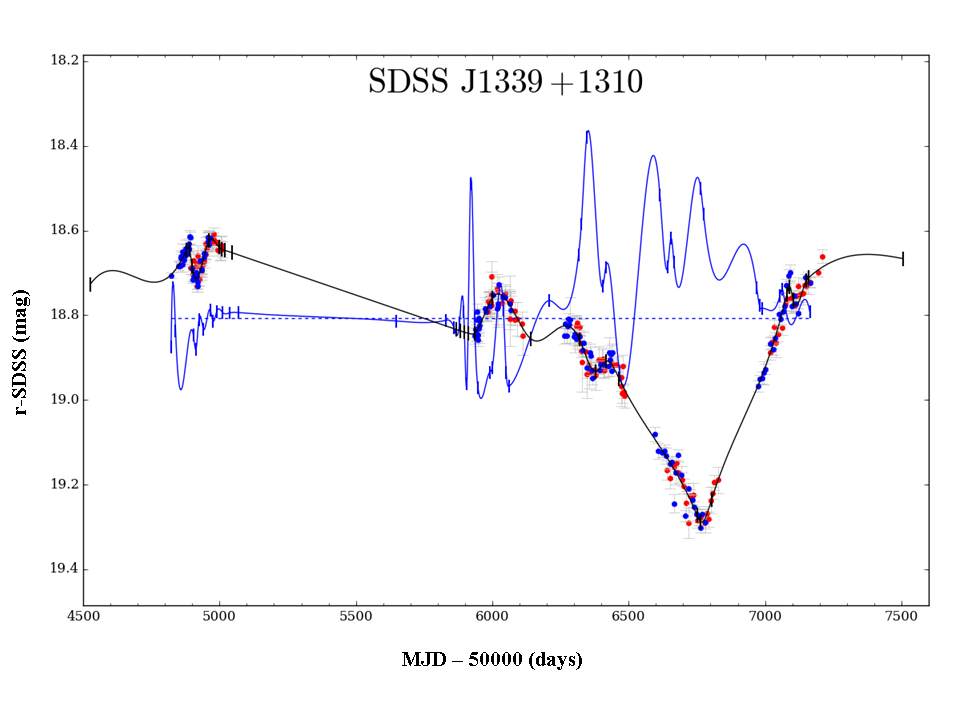}
\caption{Combined light curve in the $r$ band from the $\chi^2$ minimization with spline-like 
microlensing. The A curve (red circles), the magnitude- and time-shifted B curve (blue circles), and 
the intrinsic spline ($\eta_{\rm{int}}$ = 100 d; black line) are drawn together. The extrinsic spline 
($\eta_{\rm{ext}}$ = 50 d; blue line) describes the extrinsic variability correction that we apply to 
the curve B. The vertical ticks represent the knots.}
\label{fig:clc_splines}
\end{figure}

\begin{figure}
\centering
\includegraphics[width=7.8cm]{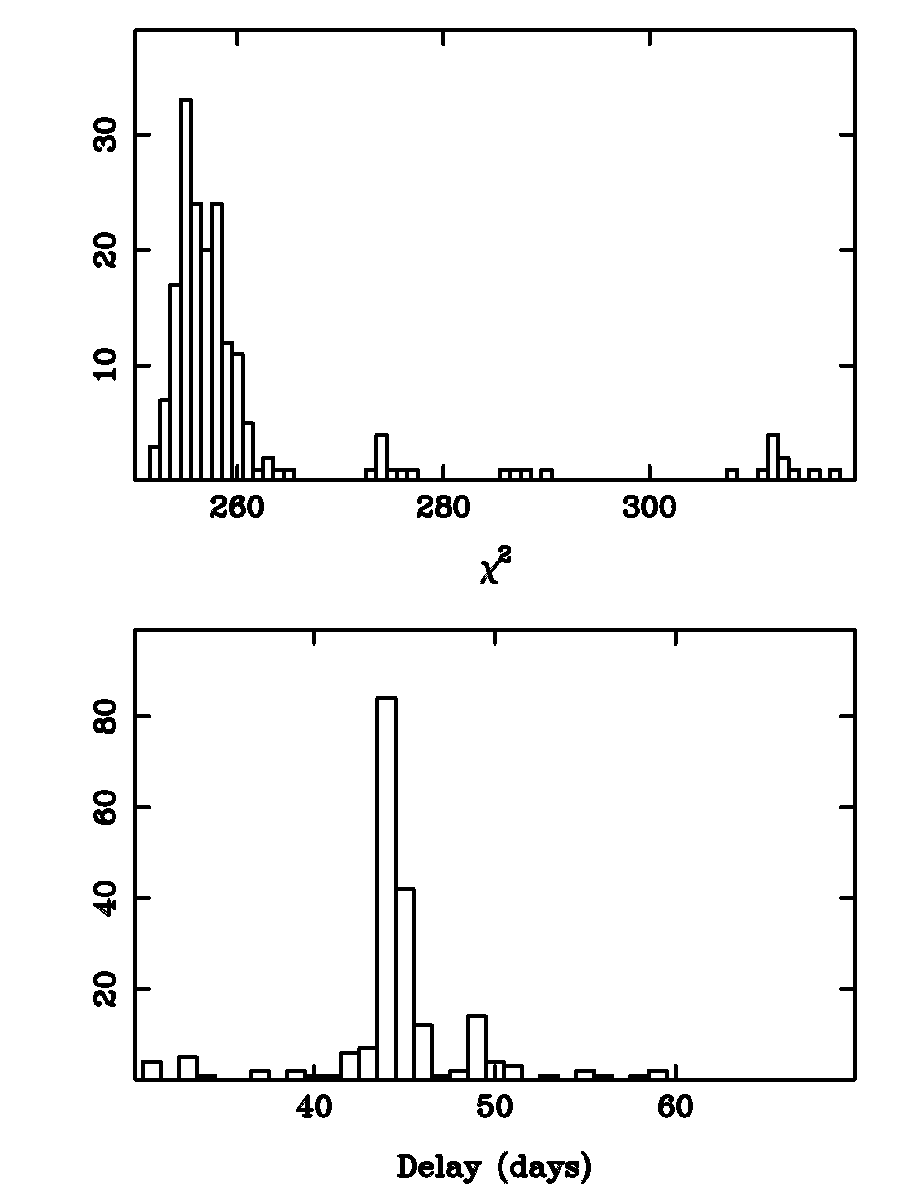}
\caption{Distributions from the $\chi^2$ minimization with spline-like microlensing. We fit the 
observed curves AB to splines with $\eta_{\rm{int}}$ = 100 d and $\eta_{\rm{ext}}$ = 50 d, starting 
from randomised initial time shifts. The top panel shows $\chi^2$ values for 200 solutions. The 
bottom panel shows the delay histogram for these solutions (see main text).}  
\label{fig:histo_spl}
\end{figure}

We obtained good fits for a knot step $\eta_{\rm{ext}}$ = 50 d, and a typical solution is shown 
in Fig.~\ref{fig:clc_splines}. The black and blue lines represent the intrinsic and extrinsic 
splines, respectively (see caption). In the top and bottom panels of Fig.~\ref{fig:histo_spl}, we 
also display $\chi^2$ and delay values for 200 solutions with different initial conditions (time 
shifts of the observed curves). The $\chi^2$ histogram is mainly concentrated within the 250$-$270 
interval, and we only considered the solutions with 250 $\leq \chi^2 \leq$ 270. These are associated 
with plausible delays ranging from 41 to 52 d. Using free-knot splines, we are dealing with a complex 
non-linear optimization involving a large set of free parameters \citep{tewes13a}, and this 
optimization sometimes yields biased results (local minima, degeneracies...). For example, our 
solutions with $\chi^2 >$ 270 generate the delay wings below 40 d and above 53 d in the bottom panel 
of Fig.~\ref{fig:histo_spl}, and thus, we removed these presumably biased estimates. The distribution 
of plausible delays has a standard deviation of 1.8 d, and it defines the intrinsic variance of the 
method. While we obtain an intrinsic uncertainty below 2 d, the total 
uncertainty (initial conditions + photometric realisations) must be larger. Unfortunately, we were 
not able to fairly assess the total error from synthetic curves as those in Sec.~\ref{sec:delseamic}. 
Such synthetic curves led to a broad delay histogram and an ambiguous distribution of $\chi^2$ 
values, which did not permit us to reject biased solutions and select the unbiased ones. The 
PyCS software was not originally designed to work with extrinsic variations more rapid than intrinsic 
ones \citep{tewes13a}, and its results seem to be less robust in that case.
  
\subsection{Adopted delay and microlensing light curve in the $r$ band}
\label{sec:delfinmicvar}

\begin{figure}
\centering
\includegraphics[width=9cm]{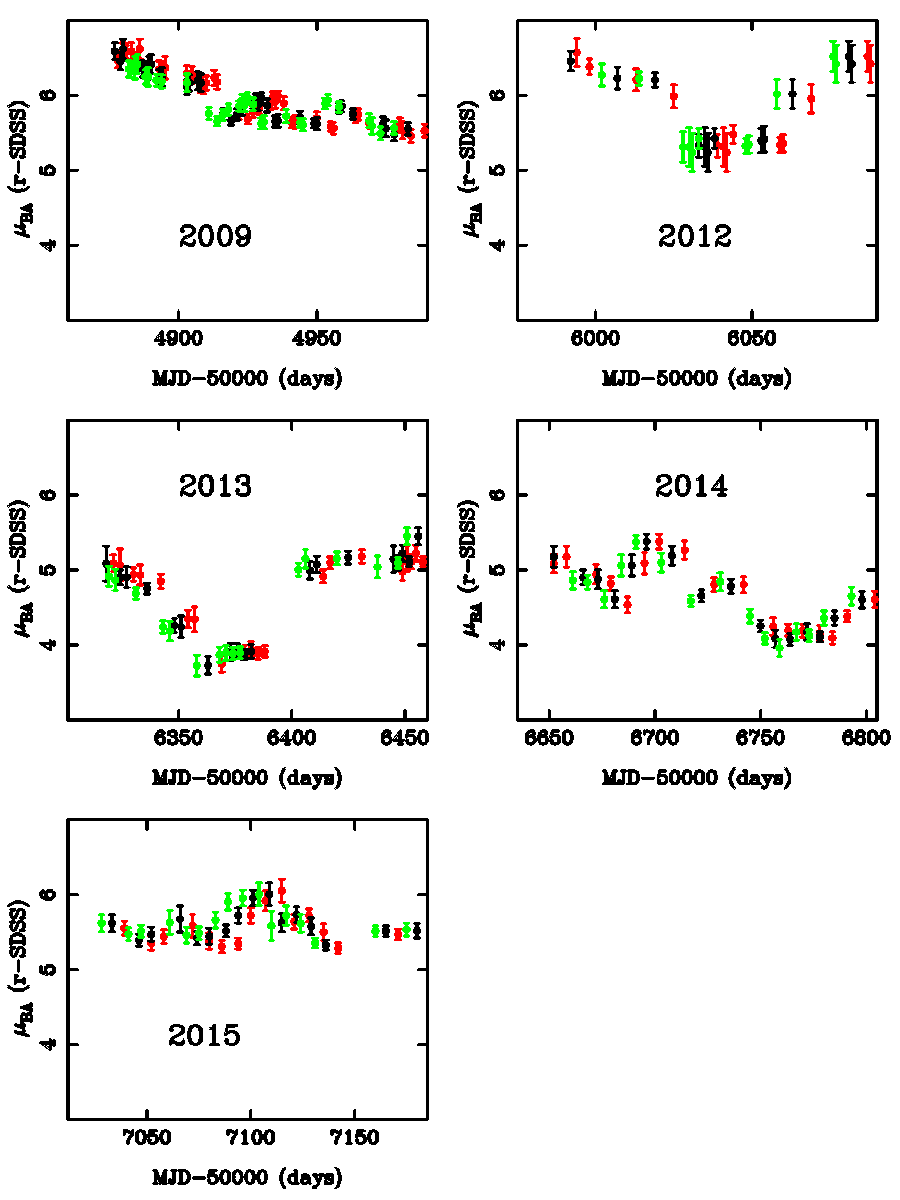}
\caption{Microlensing light curve in the $r$ band. Time evolution of the microlensing magnification 
ratio using three time delays: 41 d (red circles), 47 d (black circles) and 52 d (green circles), as 
well as 20-d bins (see main text).}  
\label{fig:miccurve}
\end{figure}

In the two previous sections, we have evaluated the time delay of \object{SDSS J1339+1310} from 
its noisy light curves with extrinsic variability. A simple microlensing model exclusively including 
seasonal variability (short-timescale microlensing events are considered to be irrelevant in the 
delay estimation and are consequently ignored) produces an 1$\sigma$ confidence interval of 46$-$49 
d. This model is reasonable because the ignored events are faster than the intrinsic variations. 
Moreover, it has a relatively small number of free parameters. From the observed brightness 
records, a second approach consisting of an extrinsic free-knot spline leads to a plausible delay 
range of 41$-$52 d. Whereas this second model incorporates microlensing variations on all 
timescales, the number of free parameters is large, and degeneracies and other problems sometimes 
bring biased delays that must be removed from delay distributions. As there is no a fair procedure to 
remove biased solutions from synthetic light curves, we cannot obtain confidence intervals in a 
standard way for the realistic (but complex) model. Instead we use the plausible delay range from the 
observed light curves as a reasonable proxy to the 1$\sigma$ error bar. This can be justified from 
simulated light curves and a modest prior on their associated delay distribution. 
We concentrated on analysing the simulated delay histogram within the range from 30 to 60 d (assuming 
the prior that shorter and longer delays are unphysical solutions), and derived a mean value and a 
standard deviation in good agreement with the plausible delay range from the observations. As the new 
delay interval includes the first (based on a simple model incorporating seasonal microlensing 
variability), we adopt $\Delta t_{\rm{AB}}$ = 47$^{+5}_{-6}$ d as our final 1$\sigma$ measurement.  

Once the time delay is measured, we can construct the microlensing light curve in the $r$ band. 
First, we made the difference light curve: $m_{\rm{A}}(t) - m_{\rm{B}}(t + \Delta t_{\rm{AB}})$, 
using 20-d bins and three delay values (41, 47 and 52 d). We then removed the macrolens and 
extinction contributions at 6225 \AA\ (through the solution at the beginning of the four paragraph in 
Sec.~\ref{sec:micline}), and converted magnitude differences into flux ratios. As we see in 
Fig.~\ref{fig:miccurve}, the microlensing curve is not very sensitive to changes in the delay 
within the range 41$-$52 d. Moreover, this curve does not critically depend on the bin size. We 
checked shorter and longer bins, and did not find any discrepancy between the corresponding curves 
and the trends depicted in Fig.~\ref{fig:miccurve}.  

\section{Conclusions}
\label{sec:end}

We presented GTC spectrophotometric data of the gravitational lens system \object{SDSS J1339+1310} in
2014. As regards previous GTC-OSIRIS observations with the R500R grism at a single epoch 
\citepalias{shalyapin14}, the exposures in 2014 were taken with the R500R and R500B grisms at two 
epochs separated by approximately the time delay of the system. The seeing and the $SNR$ were also 
better. The new spectra of the lensing galaxy G incorporate several absorption features (e.g. 
Ca\,{\sc ii} $HK$ doublet and G-band) that helped us improve our previous lens redshift 
determination. We found a practically irrelevant offset of $z_{\rm{l}}$ by a $-$0.3\%, so 
$z_{\rm{l}}$ = 0.607 $\pm$ 0.001 (1$\sigma$) from the 2014 data. The new spectra of the quasar images 
A and B include five prominent emission lines (Ly$\alpha$, Si\,{\sc iv}/O\,{\sc iv}], C\,{\sc iv}, 
C\,{\sc iii}] and Mg\,{\sc ii}), two of which were not formerly observed with the GTC. From a 
multi-component decomposition of the carbon line profiles \citep[e.g.][]{wills85,kurasz02,
marziani10,sluse11}, we derived two narrow-line flux ratios $B/A$, which were then used to achieve a 
reliable macrolens-extinction solution for a standard (linear) extinction law in G. This solution 
allowed us to remove macrolens and extinction contributions in our quasar spectra. We note that the 
narrow components of the carbon lines are related to line emitting gas with typical velocities of 
$\sim$ 700 (C\,{\sc iii}]) and 1300 (C\,{\sc iv}) km s$^{-1}$, which belong to the inner NLER 
\citep[e.g.][]{sulentic99,denney12}. 

We did not find evidence for microlensing effects on the Ly$\alpha$ and Mg\,{\sc ii} line cores. 
However, the cores of 
the Si\,{\sc iv}/O\,{\sc iv}], C\,{\sc iv} and C\,{\sc iii}] emission lines were clearly affected by 
microlensing. We obtained a microlensing magnification ratio (B relative to A) $\mu_{\rm{BA}}$ = 1.15 
(0.15 mag) for the C\,{\sc iii}] line core, as well as higher ratios of 1.3$-$1.4 ($\sim$ 0.3$-$0.4 
mag) for the high-ionization line cores (Si\,{\sc iv}/O\,{\sc iv}] and C\,{\sc iv}). In addition, 
$\mu_{\rm{BA}}$ = 1.6 ($\sim$ 0.5 mag) and $\mu_{\rm{BA}}$ = 2 (0.75 mag) for the C\,{\sc iii}] and 
C\,{\sc iv} broad-line emissions, respectively. This last ratio is about 1/3 of $\mu_{\rm{BA}} \sim$ 
5.7 ($\sim$ 1.9 mag) for the nuclear continuum at 5000 \AA. Therefore, if the size of the NCER is 
comparable to the Einstein radius of the microlensing objects \citep[stars; e.g.][]{schneider06}, 
the C\,{\sc iv} BLER should be more extended (but not much more) than the NCER. When comparing the 
C\,{\sc iv} line shape in A and B, we also detected a microlensing-induced distortion. Although the 
two blue wings coincide well with each other, there is a strong enhancement in the red wing of B 
relative to that of A. This type of distortion suggests that the C\,{\sc iv} BLER has an anisotropic 
structure \citep[e.g.][]{schneider90,abajas02,lewis04,sluse12}. For the nuclear continuum, 
$\mu_{\rm{BA}} \sim$ 5$-$6 at 4000$-$7000 \AA, with larger ratios at shorter wavelengths. All these 
spectral results (together with other time-domain results; see below) confirm that \object{SDSS 
J1339+1310} is an uncommon microlensing factory.  

We also conducted a monitoring campaign on \object{SDSS J1339+1310} in 2009, 2012$-$2015 and early 
2016 with the LT in the SDSS $r$ passband. Hence, this campaign spans five observing seasons and one 
additional month in 2016. The $r$-band light curves of the lensed quasar are characterised by typical 
photometric accuracies of 1.7\% (A) and 1.4\% (B), and show parallel V-shaped variations of A and B. 
Besides these prominent dips in both light curves, we also detected significant microlensing 
variability on different timescales, including the presence of microlensing fluctuations lasting  
50$-$100 d. This extrinsic variability should be taken into account to obtain an unbiased measurement 
of the time delay between images \citep[e.g.][]{goico98,hainline13,tewes13a}. First, 
considering only the seasonal (long-timescale) microlensing fluctuation over the period 2009$-$2015, 
we obtained a time delay of 47$^{+2}_{-1}$ d (1$\sigma$ confidence interval; A is leading). Second, 
the estimation of the delay error turned out to be difficult for a more detailed (realistic) 
microlensing model. However, using a reasonable prior on the delay distribution from simulated light 
curves (we exclusively focused on solutions in the interval 30$-$60 d), we found a 47-d value with 
$\sim$ 10\% precision. This broader range of delays practically coincides with the range of plausible 
solutions from the observed light curves and the realistic microlensing model, and we adopted 
47$^{+5}_{-6}$ d as our final 1$\sigma$ estimation. We also note that the observed interval overlaps 
with the delay interval predicted by lens models \citepalias{shalyapin14}. Finally, the time delay 
and the microlensing effects we report here are useful tools for doing several types of astrophysical 
studies \citep[e.g.][]{schneider06}.       

\begin{acknowledgements}
We wish to thank Malte Tewes for his support while using the PyCS software. We also thank the 
anonymous referee for her/his constructive criticism, which allowed us to improve the original 
manuscript. Based on observations 
made with the Gran Telescopio Canarias (GTC), installed at the Spanish Observatorio del Roque de los 
Muchachos of the Instituto de Astrof\'{\i}sica de Canarias, in the island of La Palma. This article 
is also based on observations made with the Liverpool Telescope (LT), operated on the island of La 
Palma by Liverpool John Moores University in the Spanish Observatorio del Roque de los Muchachos of 
the Instituto de Astrof\'{\i}sica de Canarias with financial support from the UK Science and 
Technology Facilities Council. We thank the staff of both telescopes for a kind interaction before, 
during and after the observations. We also used data taken from the SDSS databases, and we are 
grateful to the SDSS collaboration for doing those public databases. This research has been supported 
by the Spanish Department of Research, Development and Innovation grant AYA2013-47744-C3-2-P 
(Gravitational LENses and DArk MAtter - GLENDAMA project), and the University of Cantabria.
\end{acknowledgements}

\begin{appendix}

\clearpage

\section{Spectral extraction biases and evolution of quasar spectra}
\label{sec:specextevo}

\begin{table*}
\centering
\caption{Spectral-to-photometric flux ratios in the $r$ band.}
\begin{tabular}{cccc}
\hline\hline
\multicolumn{2}{c}{epoch 1} & \multicolumn{2}{c}{epoch 2} \\
\hline
$A1$(spec)/$A1$(phot) & $B1$(spec)/$B1$(phot) & $A2$(spec)/$A2$(phot) & $B2$(spec)/$B2$(phot) \\
\hline
0.973 $\pm$ 0.032 & 1.011 $\pm$ 0.026 & 0.992 $\pm$ 0.013 & 0.993 $\pm$ 0.010 \\                           
\hline
\end{tabular}
\label{tab:extest}
\end{table*}

\begin{figure}
\centering
\includegraphics[width=9cm]{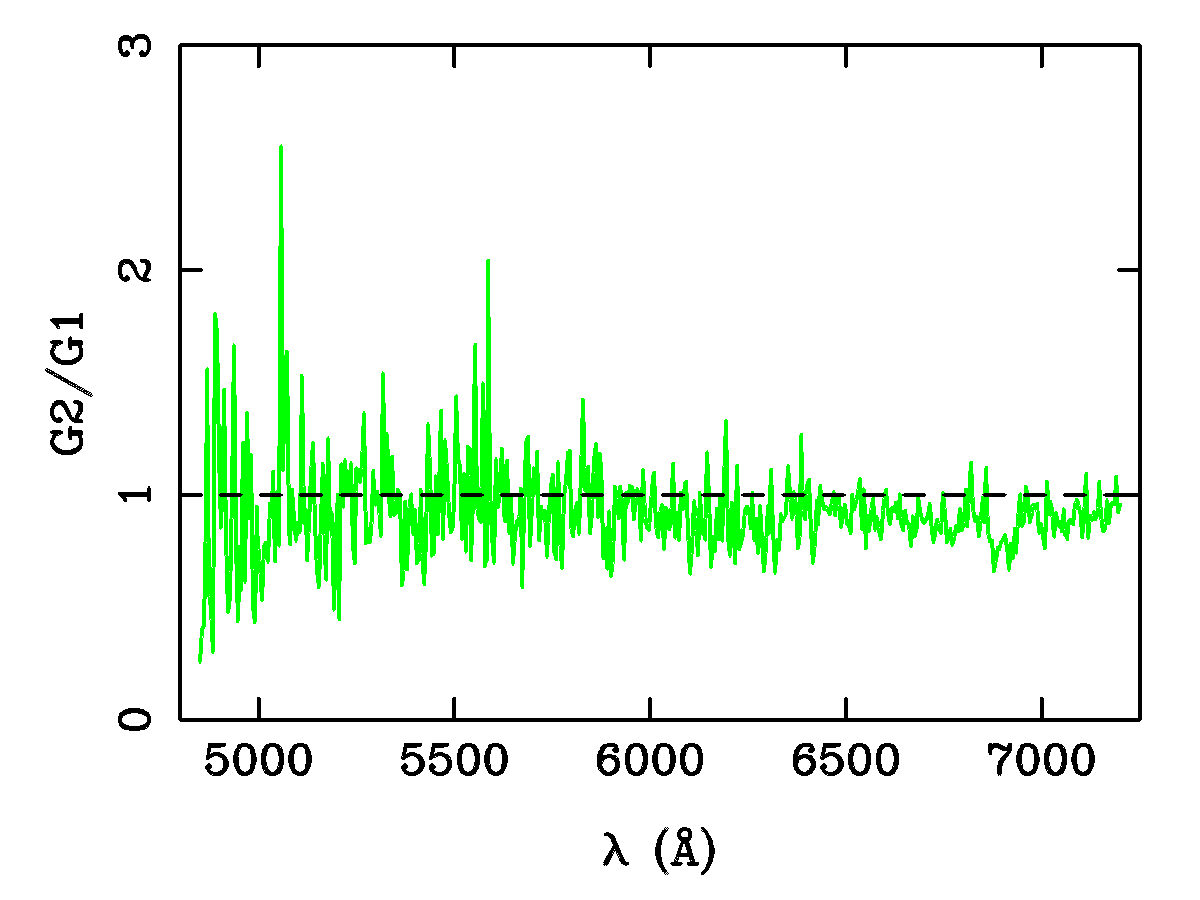}
\includegraphics[width=9cm]{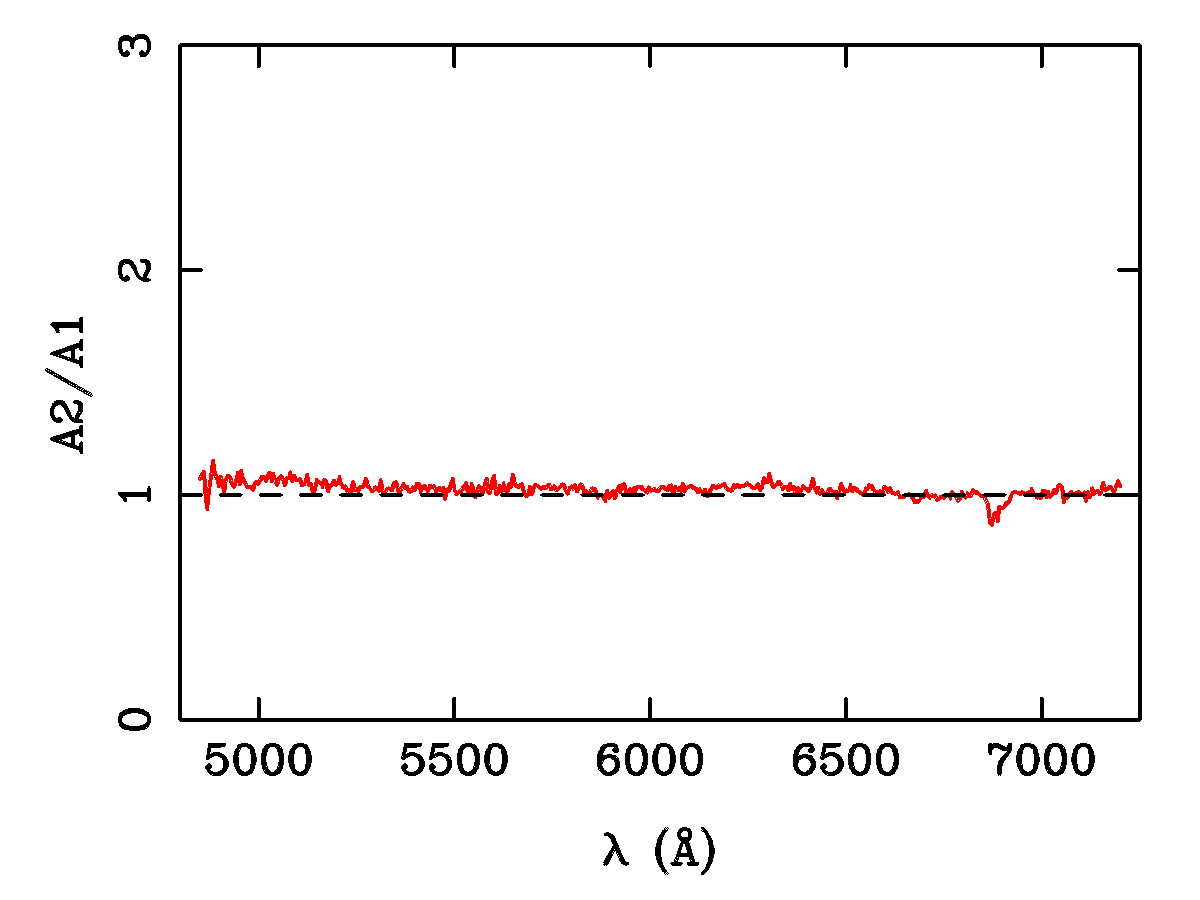}
\includegraphics[width=9cm]{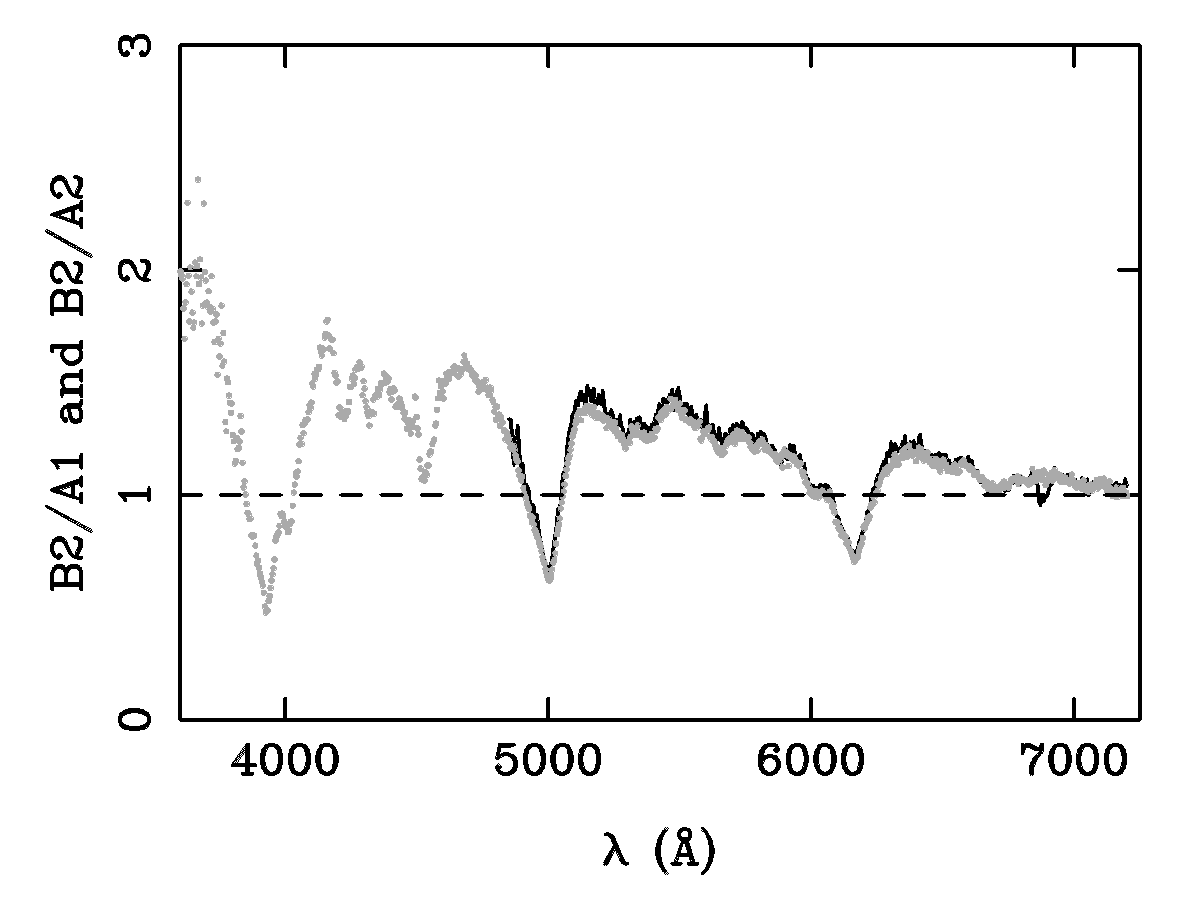}
\caption{Spectral ratios. The numbers 1 and 2 denote the first and second spectroscopy epoch. The top 
and middle panels describe the ratios for the two sources that do not show evidence of variability 
between both epochs (G and A). The bottom panel describes the ratio between the two quasar images at 
the same emission time: $B$2/$A$1 (black) and $B$2/$A$2 (grey).}  
\label{fig:specrat}
\end{figure}

\begin{figure}
\centering
\includegraphics[width=9cm]{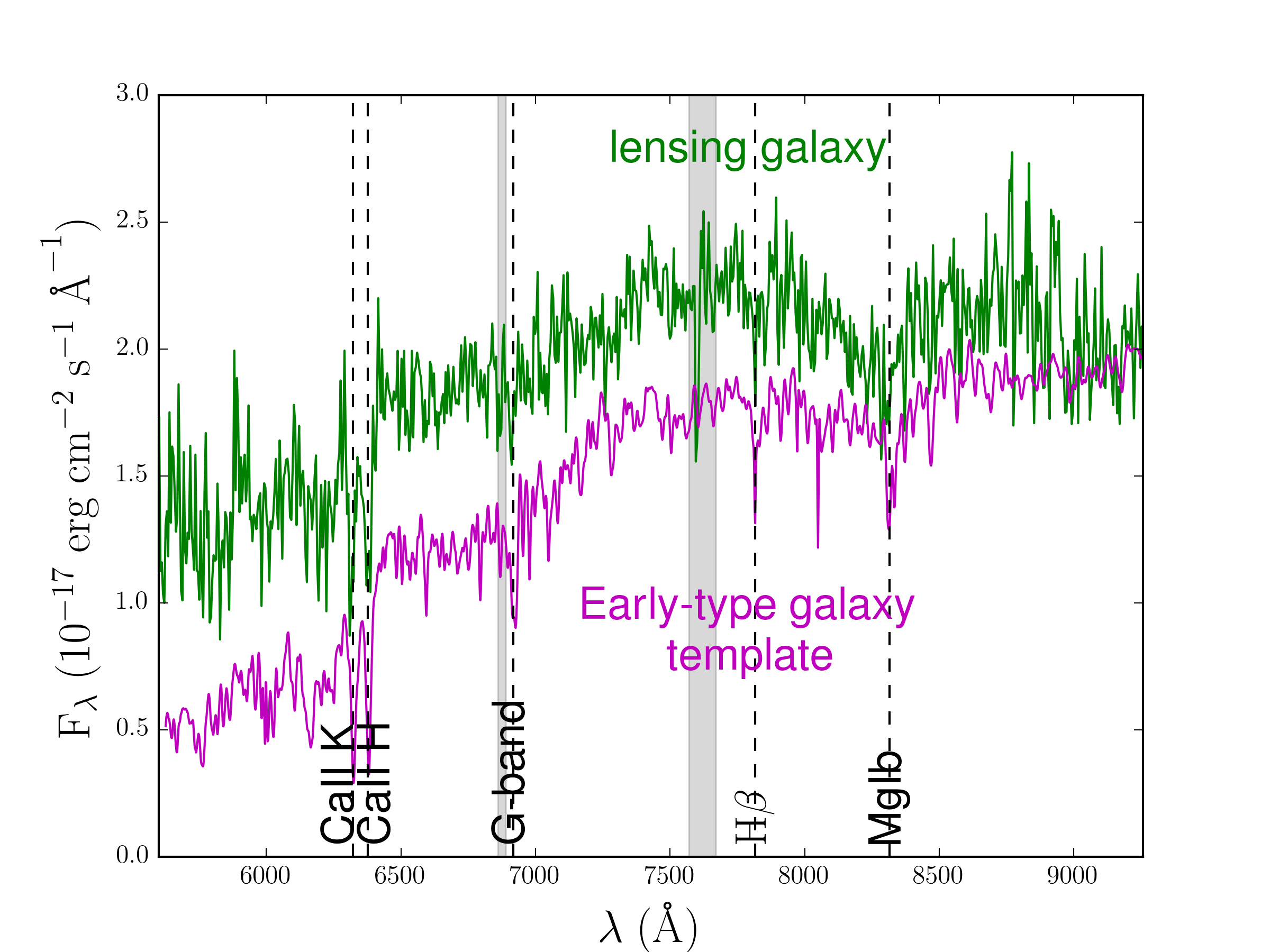}
\caption{GTC-OSIRIS-R500R spectrum of the lensing galaxy and the early-type galaxy template from the 
SDSS database. Unlike Fig.~\ref{fig:specG}, here it is included the red edge of the red grism.}
\label{fig:specredG}
\end{figure}

In Sec.~\ref{sec:spec}, we present spectra for the two quasar images (A and B) and the lensing 
galaxy G at two epochs almost separated by the time delay (see Sec.~\ref{sec:intro} and 
Sec.~\ref{sec:delmicvar}). Here, we check possible biases introduced by the spectral extraction, and 
discuss the short-term evolution of the quasar spectra.

To test the quality of the extraction technique and flux calibration, we compared quasar spectral 
fluxes averaged over the $r$ passband (using its response function) and quasar fluxes from $r$-band 
frames taken on nights close to those where we carried out spectroscopy with GTC-OSIRIS. These frames 
belong to our photometric monitoring programme with the LT (see Sec.~\ref{sec:lcur}). For the first 
spectroscopy epoch (27 March 2014), we used photometric data from LT exposures on 19 March 2014. This 
night the seeing value was $\sim$ 1\farcs5, while the seeing conditions were clearly worse on the 
closest monitoring night (26 March 2014). For the second spectroscopy epoch (20 May 2014), we 
compared the spectral fluxes with photometric data on 19 May 2014. The spectral-to-photometric flux 
ratios in the $r$ band are given in Table~\ref{tab:extest}. Taking the photometric error bars into 
account (magnitude errors), all ratios are consistent with 1. This is in good agreement with weak 
DAR-induced distorsions at $\sim$ 5500$-$7000 \AA\ (see the second paragraph in Sec.~\ref{sec:spec}), 
as well as accurate extraction and calibration procedures. In particular, there is no evidence 
supporting the existence of cross-contamination (crosstalk) between the galaxy and the quasar images, 
since such spectral extraction bias would produce ratios above or below one.

The galaxy is a non-variable faint source, and one really sees a reasonable agreement between the 
bright and light green lines in Fig.~\ref{fig:specG}. In the top panel of Fig.~\ref{fig:specrat}, we 
also show a noisy signal around one that corresponds to the ratio $G$2/$G$1 in the overlapping spectral 
region (4850$-$7200 \AA). Additionally, the LT frames indicated that the image A has the same 
$r$-band brightness (19.29 mag) at both epochs. Besides this photometric constancy, in 
Fig.~\ref{fig:specABG}, it is evident that the spectra of A at the two epochs are almost identical 
between 4850 and 7200 \AA. The spectral ratio $A$2/$A$1 confirms the lack of significant variability 
(see the middle panel of Fig.~\ref{fig:specrat}), since its averaged deviation from one is only 2.5\%,
which does not exceed the typical uncertainty in $A$2/$A$1 in the $r$ band of 2.8\% as estimated from 
the spectral-to-photometric flux ratios in Table~\ref{tab:extest}. Therefore, one can use the second 
spectrum of A instead the first when estimating delay-corrected flux ratios $B/A$. This approach has 
a key advantage over the direct one (based on a comparison between B at the epoch 2 and A at the 
epoch 1): it provides plausible information over a broader range of wavelengths (3600$-$7200 \AA), 
including the Ly$\alpha$, Si\,{\sc iv}/O\,{\sc iv}], C\,{\sc iv} and C\,{\sc iii}] emission lines. 

Unlike G and A, the B image substantially varies on a timescale of $\sim$ 50 d (see the bright and 
light blue lines in Fig.~\ref{fig:specABG}). However, in the absence of dust extinction and chromatic
microlensing, the spectral ratios $B$2/$A$1 and $B$2/$A$2 must be practically constant for all 
wavelengths. In the bottom panel of Fig.~\ref{fig:specrat}, these ratios unambiguously prove the 
existence of differential extinction-microlensing effects. We remark that the measured signal 
(incorporating strong dips in the emission line regions) is a fair indicator of the presence of 
microlensing. 

It is also worth mentioning that the red grism data allow us to estimate the flux 
ratios for the C\,{\sc iii}] and Mg\,{\sc ii} line cores at the first epoch. The average variation of
the flux of A in the $r$ band on a timescale of $\sim$ 50 d is only a few percent (see 
Fig.~\ref{fig:clc_step}). Hence, typical intrinsic variations of line cores on this timescale are 
expected to be less than or similar to 1\%, and consequently, smaller than the relative uncertainties 
in the line-core flux ratios from the red grism data ($\sim$ 5$-$10\%; see 
Appendix~\ref{sec:linec}). In other words, the single-epoch ratios for the C\,{\sc iii}] and Mg\,{\sc 
ii} cores are most likely unbiased by intrinsic variability. In addition, despite the global 
parallelism between the spectra of G and the early-type galaxy template in Fig.~\ref{fig:specG}, 
there is a relative drop in the flux of G at $\sim$ 9000 \AA\ (see Fig.~\ref{fig:specredG}). Although 
this feature could be true, it may be also an artefact. However, even if G is progressively 
contaminating the spectrum of B at the reddest wavelengths, the contamination by galaxy light would 
play the role of an additive pseudo-continuum and would not seriously affect the estimation of flux 
ratios for the Mg\,{\sc ii} line.

\clearpage
             
\section{Line-core flux ratios}
\label{sec:linec}

\begin{table*}
\centering
\caption{Line-core flux ratios.}
\begin{tabular}{lcccccc}
\hline\hline
Line core & Central $\lambda$ (\AA) & \multicolumn{2}{c}{$(B/A)_{\rm{core}}$} & 
$\Delta m_{\rm{core}}$ (mag) & Degree of ionization & $E_{\rm{exc}}$ or $E_{\rm{ion}}$ (eV) \\
\hline
 & & Paper I & This work & & & \\
\hline
Ly$\alpha$ & 3930 & &                    0.303 $\pm$ 0.010 & $-$1.30 $\pm$ 0.04 & High & 10.2 \\
Si\,{\sc iv}/O\,{\sc iv}] & 4525 & &     0.356 $\pm$ 0.019 & $-$1.12 $\pm$ 0.06 & High & 58.0 \\
C\,{\sc iv} & 5005 & 0.39 $\pm$ 0.03 &   0.365 $\pm$ 0.007 & $-$1.09 $\pm$ 0.02 & High & 83.5 \\           
C\,{\sc iii}] & 6165 & 0.27 $\pm$ 0.03 & 0.275 $\pm$ 0.007 & $-$1.40 $\pm$ 0.03 & Low  & 35.6 \\          
Mg\,{\sc ii} & 9047 & 0.24 $\pm$ 0.03 &  0.234 $\pm$ 0.019 & $-$1.58 $\pm$ 0.09 & Low  &  7.6 \\                           
\hline
\end{tabular}
\label{tab:lcfrat}
\end{table*}

\begin{figure}
\centering
\includegraphics[width=9cm]{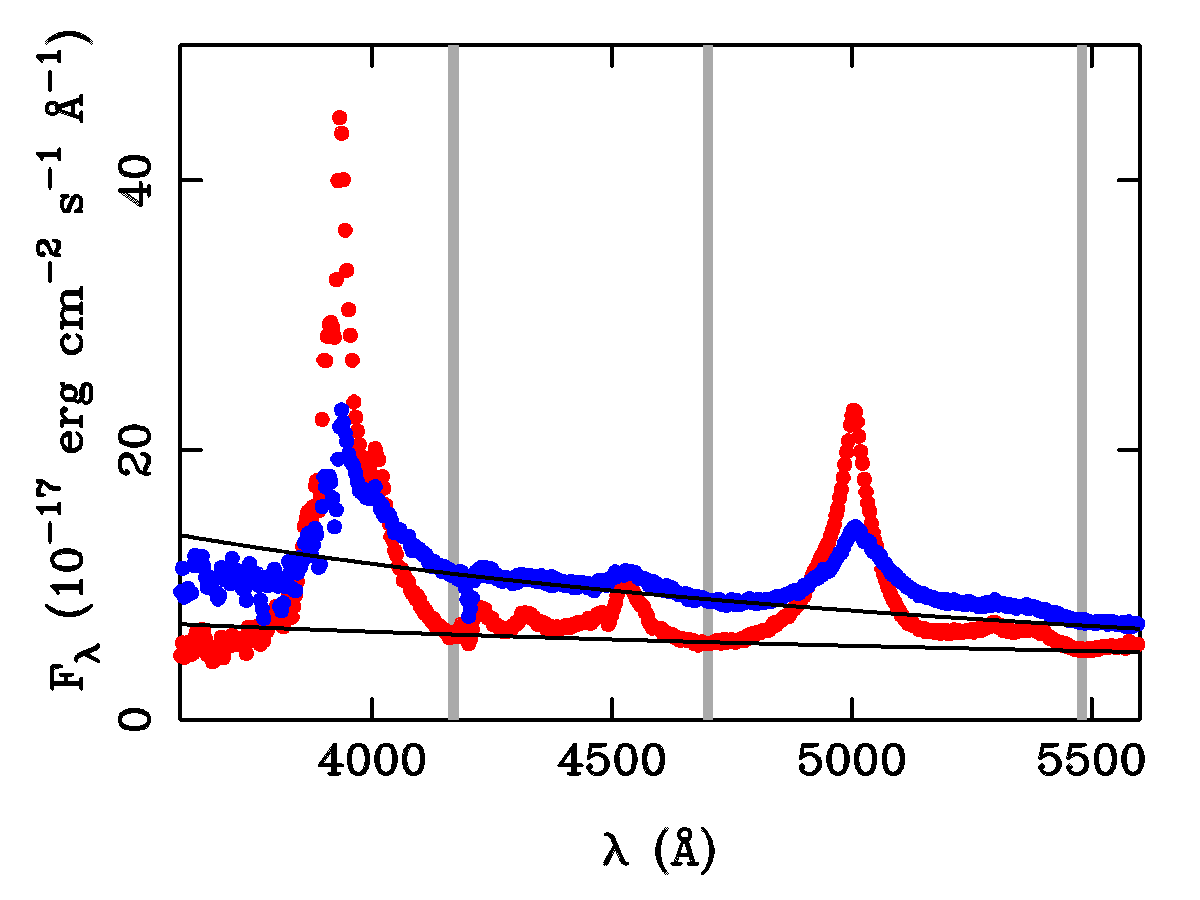}
\caption{Power-law fits to the nuclear continuum in the blue grism spectra of the quasar. The red and 
blue circles represent the spectral profiles of A and B, respectively. We show the three pure 
continuum windows close to the Ly$\alpha$, Si\,{\sc iv}/O\,{\sc iv}] and C\,{\sc iv} lines (grey 
highlighted regions), together with the fitted power laws (black lines).}
\label{fig:PLfitcont}
\end{figure}

\begin{figure}
\centering
\includegraphics[width=9cm]{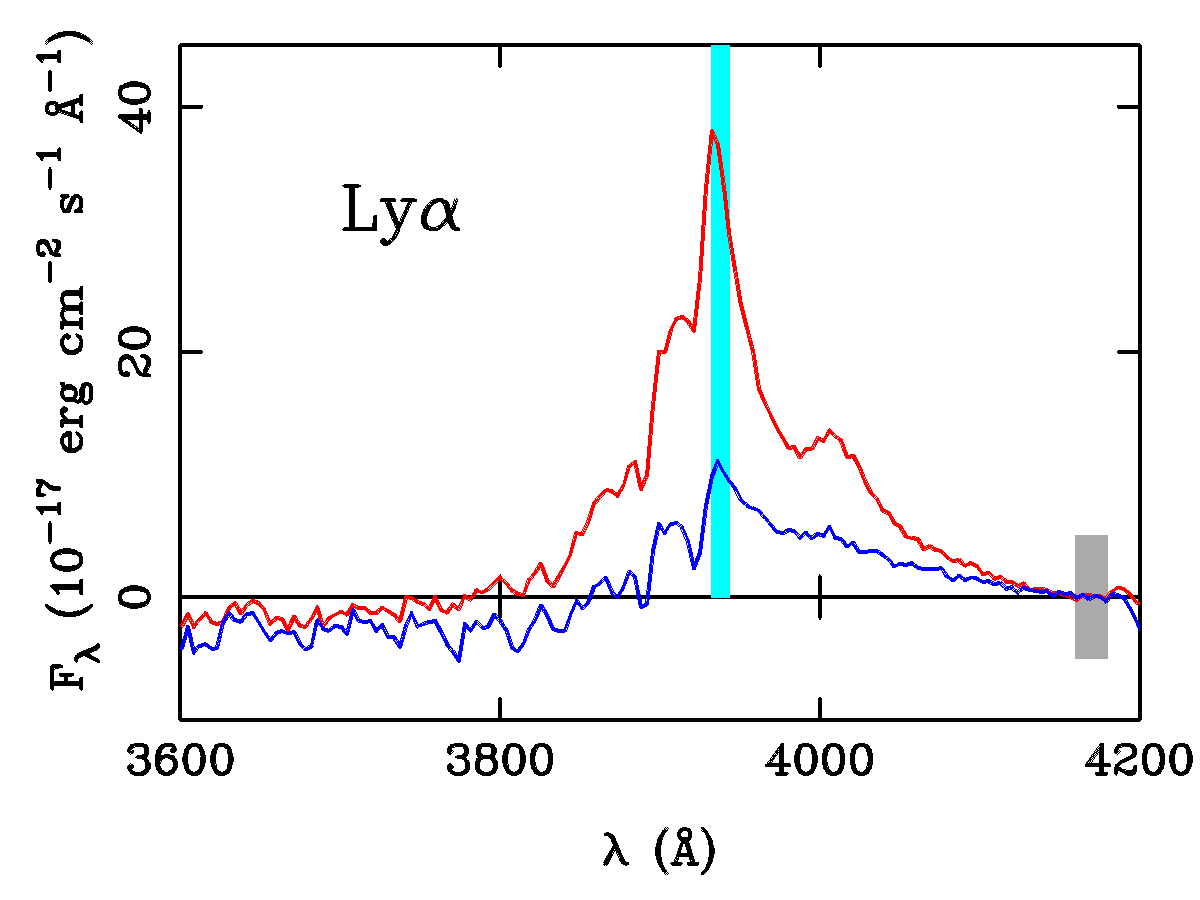}
\caption{Ly$\alpha$ emission line profiles. The red and blue lines describe the profiles for the 
images A and B, respectively. We also highlight the pure continuum window closest to the Ly$\alpha$ 
emission (grey) and the red side of the line core, i.e. the 15 \AA\ wide region to the right of the 
main Ly$\alpha$ peak (cyan).}
\label{fig:Lya}
\end{figure}

\begin{figure}
\centering
\includegraphics[width=9cm]{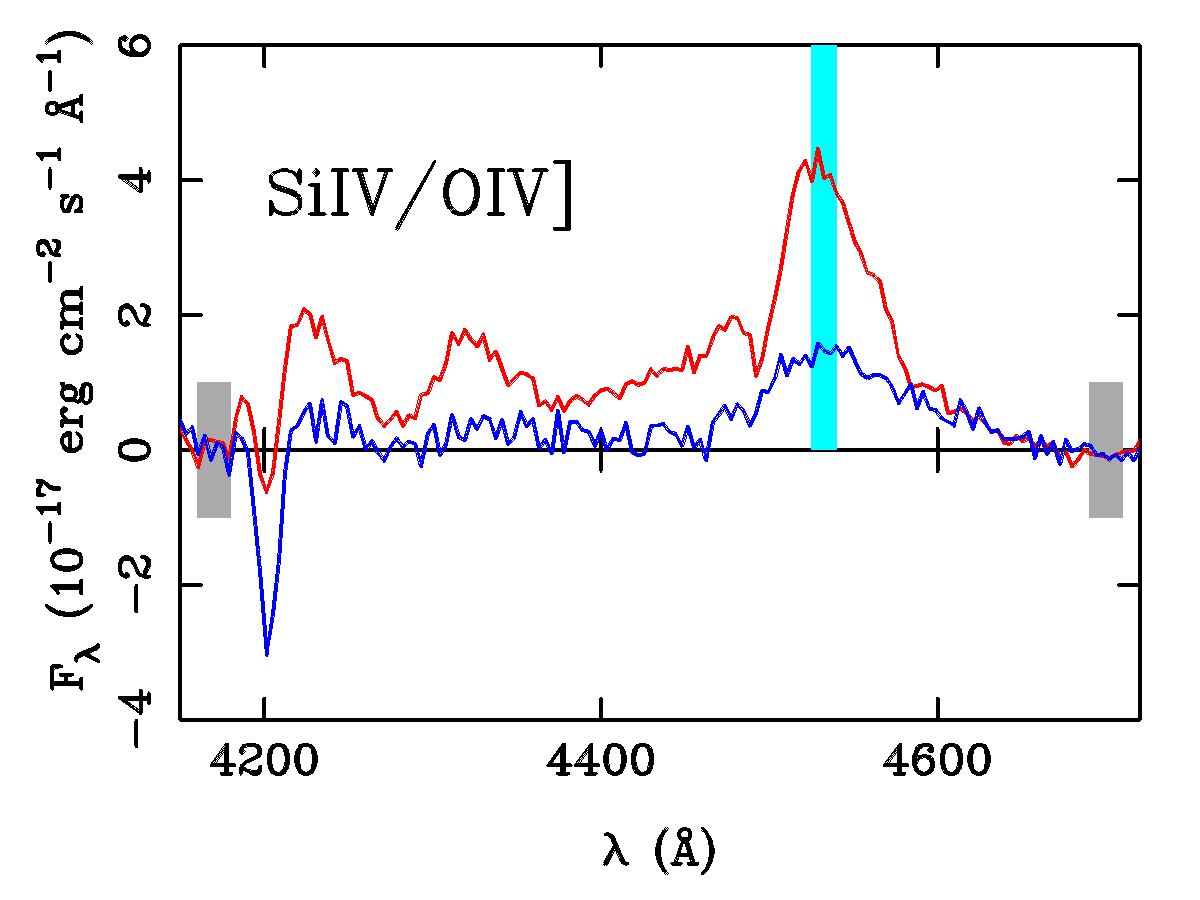}
\includegraphics[width=9cm]{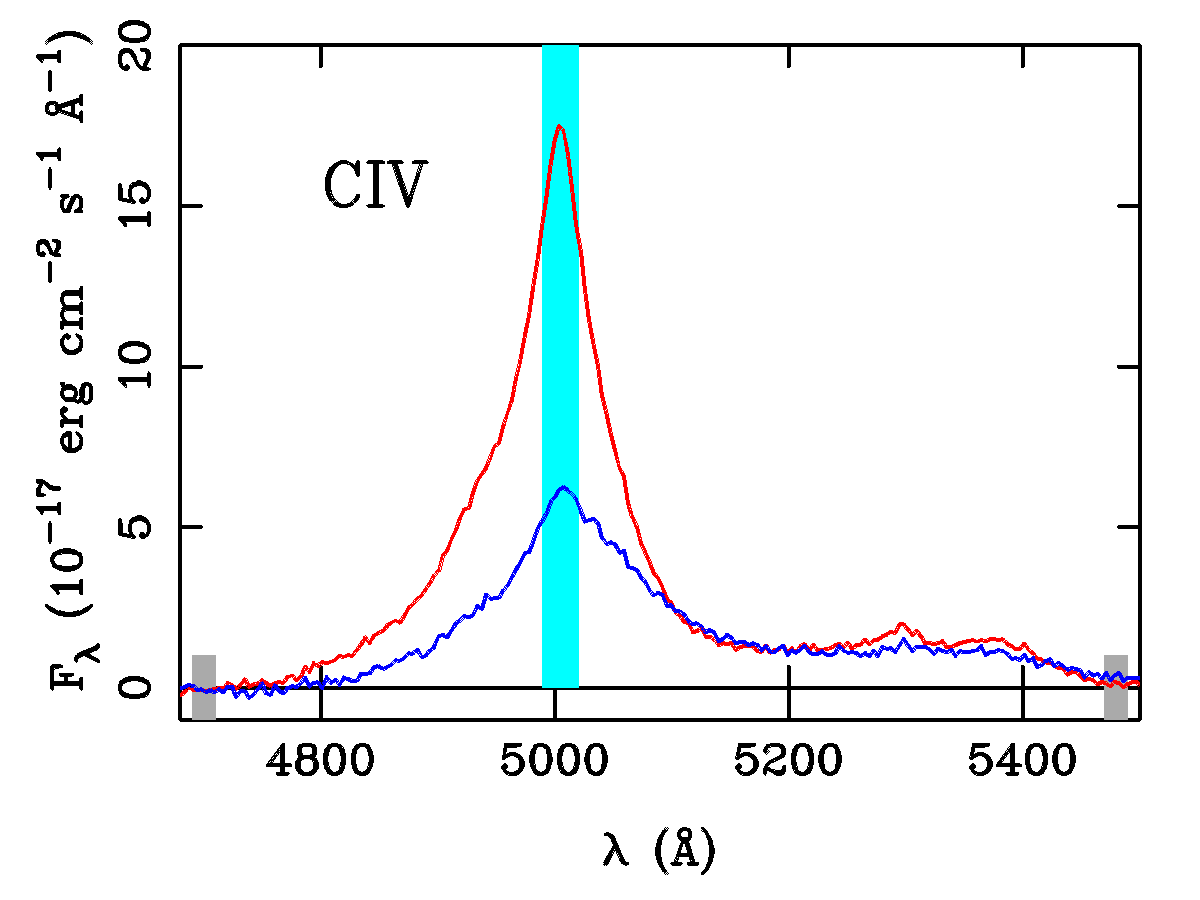}
\caption{Si\,{\sc iv}/O\,{\sc iv}] and C\,{\sc iv} emission line profiles. The red and blue lines 
trace the profiles for the images A and B, respectively. The grey highlighted regions represent some 
pure continuum windows (see main text), whereas the red side of the Si\,{\sc iv}/O\,{\sc iv}] line 
core (top panel) and the whole core of the C\,{\sc iv} line (bottom panel) are marked in cyan.}
\label{fig:SiOCiv}
\end{figure}

\begin{figure}
\centering
\includegraphics[width=9cm]{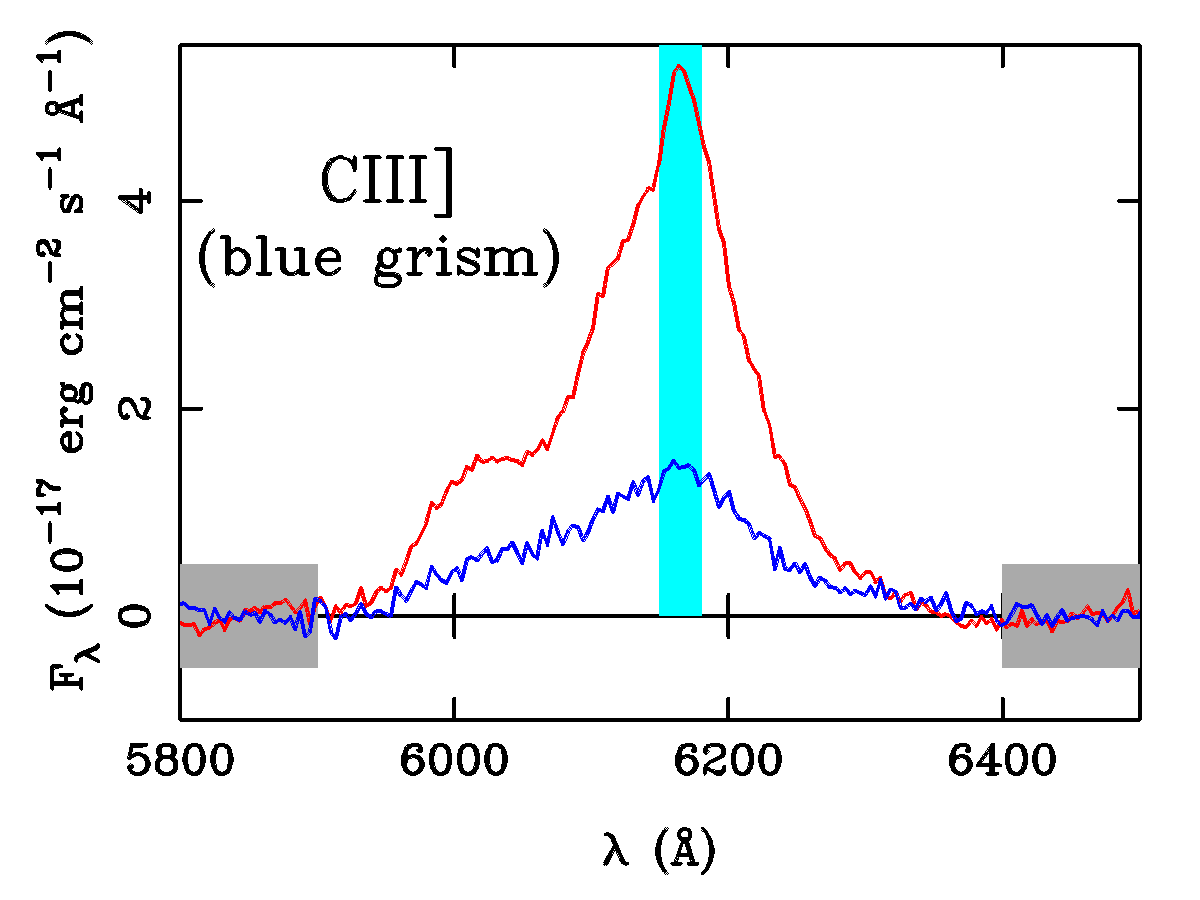}
\includegraphics[width=9cm]{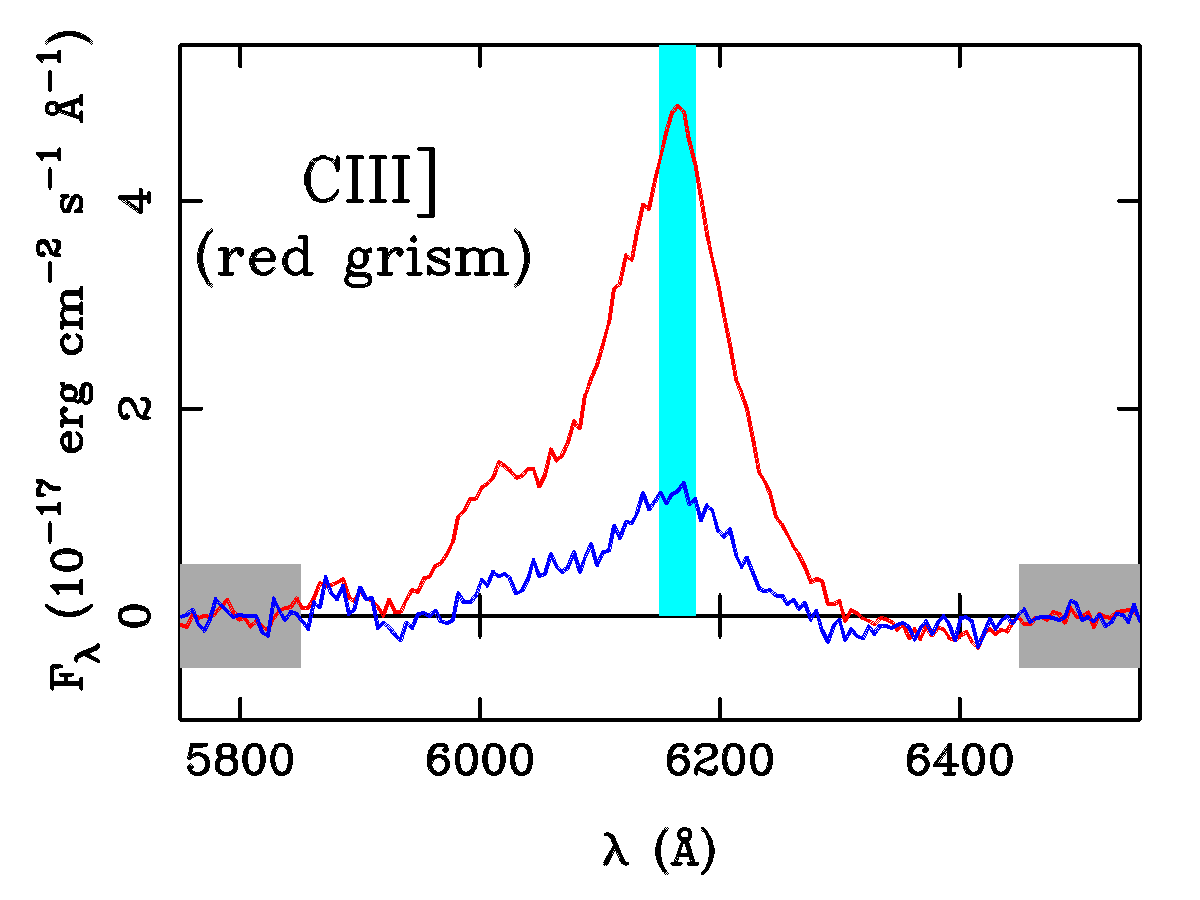}
\caption{C\,{\sc iii}] emission line profiles from the blue and red grism spectra. The red and blue 
lines describe the profiles for the images A and B, respectively. The grey rectangles are 
associated with total continuum regions covering 100 \AA, and the C\,{\sc iii}] line core is marked 
in cyan.}
\label{fig:Ciii}
\end{figure}

\begin{figure}
\centering
\includegraphics[width=9cm]{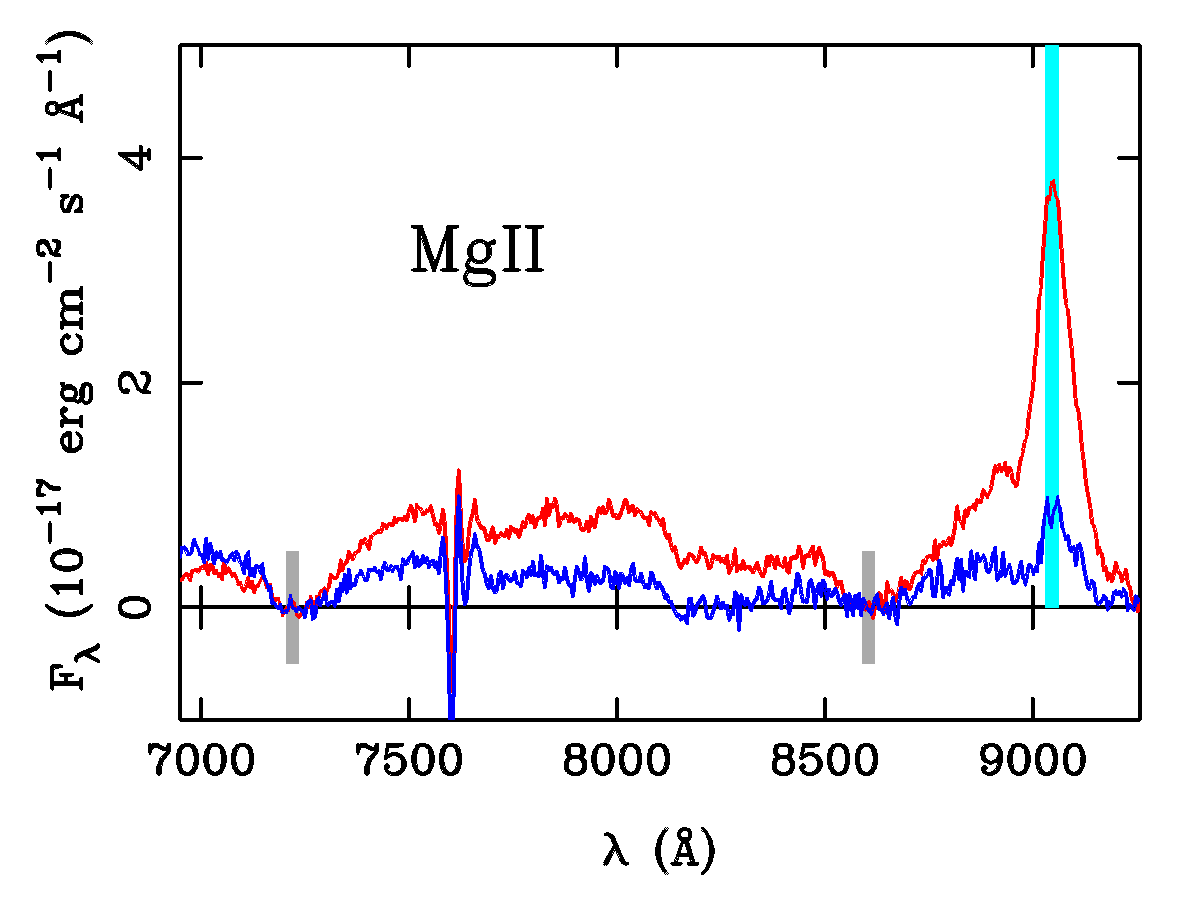}
\caption{Mg\,{\sc ii} emission line profiles. The red and blue lines describe the profiles for the 
images A and B, respectively. We highlight three 30 \AA\ wide regions: two (total) continuum windows 
to the left of the Mg\,{\sc ii} line (grey) and the Mg\,{\sc ii} line core (cyan).}
\label{fig:Mgii}
\end{figure}

First, we used a single power law to fit the nuclear continuum in the blue grism spectra of A and B 
at rest-frame wavelengths $\leq$ 2200 \AA. According to \citet{vandenberk01} and \citet{kurasz02}, we 
took four pure (nuclear) continuum windows to perform the two fits: C1 $\equiv$ 4160$-$4180 \AA, C2 
$\equiv$ 4690$-$4710 \AA, C3 $\equiv$ 5470$-$5490 \AA, and C4 $\equiv$ 6460$-$6480 \AA\ (see 
Fig.~\ref{fig:PLfitcont}). These windows are around 1290 \AA\ (C1), 1455 \AA\ (C2), 1695 \AA\ (C3) 
and 2000 \AA\ (C4) in the quasar rest frame. However, the total continuum under some line cores may 
differ from the power-law one. Even though the Balmer continuum does not significantly contaminate 
the quasar spectra at rest-frame wavelengths $\leq$ 2200 \AA, the Fe pseudo-continuum (iron forest) 
plays a role, and this is not taken into account through a global power-law fit 
\citep[e.g.][]{vestergaard01}.

For the blue grism spectra, in a second step, we considered the contamination of its four line cores 
by iron emission \citep[e.g.][]{vandenberk01,vestergaard01}:
\begin{description}
\item[{\bf Ly$\alpha$}] We assumed that the Fe contribution to the Ly$\alpha$ line profile is 
negligibly small. However, the blue side of this emission line is noticeably absorbed by clouds of 
neutral hydrogen (see Fig.~\ref{fig:PLfitcont}) and it is not considered here. We exclusively focused 
on the red side of the line core.
\item[{\bf Si\,{\sc iv}/O\,{\sc iv}]}] The red side of the Si\,{\sc iv}/O\,{\sc iv}] line core is not 
contaminated by the iron forest.
\item[{\bf C\,{\sc iv}}] The C\,{\sc iv} line core is basically uncontaminated by Fe emission. 
\item[{\bf C\,{\sc iii}]}] The C\,{\sc iii}] line lies above an iron (Fe\,{\sc iii}) complex. Hence, 
in this study, we did not use the global power-law fits to estimate the continuum under the C\,{\sc 
iii}] line core. Instead we obtained local polynomial-law fits, defining 100 \AA\ wide continuum 
windows close to the red and blue side of the emission line.
\end{description}

Third, in analysing the red grism spectra, we also used local polynomial-law fits to determine the 
total continuum under the C\,{\sc iii}] line core. The C\,{\sc iv} line is located on the blue edge 
of the red grism (see the bright red and blue lines in Fig.~\ref{fig:specABG}), and it was not 
further considered. The Mg\,{\sc ii} line at rest-frame wavelengths $\sim$ 2800 \AA\ lies above the 
3000 \AA\ bump, which is a very broad feature from $\sim$ 2200 to 4000 \AA\ 
\citep[e.g.][]{vandenberk01}. This mostly consists of blends of Fe\,{\sc ii} line emission and 
Balmer continuum emission \citep[e.g.][]{wills85}. Here, we carried out local linear fits to 
minimise the Fe\,{\sc ii}-Balmer contamination. Unfortunately, the Mg\,{\sc ii} emission line appears 
close to the red edge of the red grism, so we cannot define continuum windows to the right of its red 
side. Details on each emission line core, as well as its corresponding flux ratio, are given in the 
following four paragraphs. 

\subsection{Ly$\alpha$}

After subtracting the power-law curves that best fit the nuclear continuum of A and B, we derived the 
Ly$\alpha$ line profiles in Fig.~\ref{fig:Lya}. The grey rectangle at $\lambda \sim$ 4170 \AA\ 
denotes one out of the four pure continuum windows that we used (C1), whereas the cyan highlighted 
region indicates the red side of the line core (see above). The line core has a width of 30 \AA\ (see 
\citetalias{shalyapin14}), but its red side is only 15 \AA\ wide (3930$-$3945 \AA). We found a first 
flux ratio $(B/A)_{\rm{red-core}}$ = 0.294. The red core in Fig.~\ref{fig:Lya} was then slightly 
shifted towards the interval 3935$-$3950 \AA\ (to try to avoid possible absorption by neutral 
hydrogen in the vicinity of the BLER and the NLER), leading to a second estimate $(B/A)_{\rm{red-core}}$ = 
0.312. After accounting for the signal noise ($\sigma_{\rm{noise}}$ = 0.003), the error in the 
power-law fits to the continuum of A and B ($\sigma_{\rm{cont}}$ = 0.001), and the uncertainty in the
choice of the red core ($\sigma_{\rm{red}}$ = 0.009), our final 1$\sigma$ measurement was 
$(B/A)_{\rm{core}}$ = 0.303 $\pm$ 0.010 (see Table~\ref{tab:lcfrat}). The flux ratio for the red core  
was reasonably assumed as a reliable tracer of the flux ratio for the whole core, and the three 
uncertainties were added in quadrature. In Fig.~\ref{fig:Lya}, we can appreciate the presence of 
N\,{\sc v} emission at about 4000 \AA. This increases the fluxes of the Ly$\alpha$ red core in $\sim$ 
2\% (B) and 1\% (A), and the two increments partially cancel each other when computing the flux 
ratio. More specifically, our estimate of $B/A$ only deviates by $\sim$ 1\% from the uncontaminated 
flux ratio, so the expected deviation of $\sim$ 0.003 is well below the adopted uncertainty and is 
irrelevant in subsequent analyses.  

\subsection{Si\,{\sc iv}/O\,{\sc iv}] and C\,{\sc iv}}

We used the global power-law fits to find the continuum levels in the spectral regions of the 
Si\,{\sc iv}/O\,{\sc iv}] and C\,{\sc iv} emissions, and then obtained the corresponding 
continuum-subtracted profiles. For the Si\,{\sc iv}/O\,{\sc iv}] line, its profiles appear in the top 
panel of Fig.~\ref{fig:SiOCiv}, where the two grey highlighted regions indicate the pure continuum 
windows C1 and C2, and the red side of the line core (4525$-$4540 \AA) is marked in cyan. Taking the 
signal noise and the error in the power-law fits into account, we measured $(B/A)_{\rm{core}}$ = 
$(B/A)_{\rm{red-core}}$ = 0.356 $\pm$ 0.019 (see Table~\ref{tab:lcfrat}). The final 1$\sigma$ 
uncertainty is dominated by $\sigma_{\rm{noise}}$ = 0.017. In the bottom panel of 
Fig.~\ref{fig:SiOCiv}, we show the C\,{\sc iv} line profiles, remarking the core of the C\,{\sc iv} 
emission at 4990$-$5020 \AA\ with cyan colour. The pure continuum windows C2 and C3 are also marked 
with two grey rectangles. The flux ratio for this core was 0.365 $\pm$ 0.007 (see 
Table~\ref{tab:lcfrat}). As usual, we added in quadrature the uncertainties coming from the signal 
noise ($\sigma_{\rm{noise}}$ = 0.006) and the continuum subtraction ($\sigma_{\rm{cont}}$ = 0.004). 

\subsection{C\,{\sc iii}]}

C\,{\sc iii}] emission is apparent in the spectra from both grisms. Therefore, we measured two 
independent flux ratios for the C\,{\sc iii}] core. For each pair of spectra (grism), in the region 
of this emission line, we subtracted linear and quadratic interpolations of the continuum of A and B. 
These interpolations were based on fluxes within total continuum regions defined by two 100 \AA\ wide 
windows, one to the left of the line and the other to its right. We set the continuum regions CC1 
$\equiv$ [5800$-$5900 \AA, 6400$-$6500 \AA], CC2 $\equiv$ [5750$-$5850 \AA, 6450$-$6550 \AA] and CC3 
$\equiv$ [5830$-$5930 \AA, 6350$-$6450 \AA]. In the top panel of Fig.~\ref{fig:Ciii}, we display line 
profiles from the blue grism spectra, using the CC1 region and a linear fit. Line profiles from the 
red grism spectra (using the CC2 region and a quadratic fit) are also depicted in the bottom panel of 
Fig.~\ref{fig:Ciii}. The total continuum windows and the line core at 6150$-$6180 \AA\ are marked in 
grey and cyan, respectively. Considering the six possible combinations between continuum region and 
continuum shape, and comparing the results, we obtained the 1$\sigma$ intervals $(B/A)_{\rm{core}}$ = 
0.282 $\pm$ 0.009 (blue grism) and $(B/A)_{\rm{core}}$ = 0.260 $\pm$ 0.013 (red grism). We then 
combined these two intervals to yield a final measurement $(B/A)_{\rm{core}}$ = 0.275 $\pm$ 0.007 
(see Table~\ref{tab:lcfrat}). A flux ratio to the 2$-$3\% is achieved for the C\,{\sc iii}] core, and 
the new uncertainty is about four times lower than the error quoted in \citetalias{shalyapin14}. 

\subsection{Mg\,{\sc ii}}

In order to analyse the Mg\,{\sc ii} emission line, we took two total continuum windows at 
7205$-$7235 \AA\ and 8590$-$8620 \AA, and then performed linear fits. These fits were used to remove 
the continuum levels in the region of the Mg\,{\sc ii} line, and the associated line profiles are 
shown in Fig.~\ref{fig:Mgii}. In this Fig.~\ref{fig:Mgii}, the first two grey rectangles denote the 
two continuum windows, whereas the region in cyan is highlighting the Mg\,{\sc ii} core at 
9032$-$9062 \AA. Adding in quadrature the uncertainties coming from the signal noise 
($\sigma_{\rm{noise}}$ = 0.005) and the continuum subtraction ($\sigma_{\rm{cont}}$ = 0.018), we 
inferred the 1$\sigma$ interval $(B/A)_{\rm{core}}$ = 0.234 $\pm$ 0.019 (see Table~\ref{tab:lcfrat}).

Apart from the new and previous 1$\sigma$ measurements of the line-core flux ratios, 
Table~\ref{tab:lcfrat} also provides the new magnitude differences $\Delta m_{\rm{core}} = 2.5 \log 
(B/A)_{\rm{core}}$ and their errors (fifth column), as well as relevant information on the degree of 
ionization of the emitting gas \citep[sixth column; e.g.][and references 
therein]{krolik91,gaskell07} and the total energy that is required to produce each emitting excited 
or ionised atom (i.e. $E_{\rm{exc}}$ or $E_{\rm{ion}}$ in the seventh column)\footnote{See the NIST 
database at \url{http://physics.nist.gov/PhysRefData/ASD/ionEnergy.html}}. For the Si\,{\sc 
iv}/O\,{\sc iv}] emission, we only considered the energies for the processes Si\,{\sc i} 
$\rightarrow$ Si\,{\sc ii} $\rightarrow$ Si\,{\sc iii} $\rightarrow$ Si\,{\sc iv}. Taking 
$E_{\rm{ion}}$ = 103.7 eV for O\,{\sc iv}, this would lead to an average energy of 80.9 eV, which is 
slightly larger than $E_{\rm{ion}} \sim$ 60 eV for Si\,{\sc iv}. The information in the last two 
columns of Table~\ref{tab:lcfrat} is used in Sec.~\ref{sec:miclinec}.

\end{appendix}

\end{document}